\documentclass[usegraphicx,usenatbib]{mn2e}

\title[How does galaxy environment matter?]{How does galaxy environment matter? The relationship between galaxy environments, colour and stellar mass at $0.4 < z < 1$ in the Palomar/DEEP2 survey}

\author[Gr\" utzbauch et al.]{Ruth Gr\"utzbauch$^{1}$\thanks{email: ruth.grutzbauch@nottingham.ac.uk}, 
Christopher J. Conselice$^{1}$, Jes\'us Varela$^{2,3}$, Kevin Bundy$^{4}$, \newauthor Michael C. Cooper$^{5}$, Ramin Skibba$^{6}$,
Christopher N. A. Willmer$^{6}$ \\
$^{1}$ School of Physics and Astronomy, University of Nottingham, UK \\
$^{2}$ Instituto de Astrof\'isica de Canarias (IAC), E-38200 La Laguna, Tenerife, Spain\\
$^{3}$  Depto. Astrof\'isica, Universidad de La Laguna (ULL), E-38206 La Laguna, Tenerife, Spain\\
$^{4}$ Department of Astronomy, University of California at Berkeley, USA\\
$^{5}$ Spitzer Fellow, Steward Observatory, University of Arizona, USA \\
$^{6}$ Steward Observatory, University of Arizona, USA \\
}

\begin{document}

\date{Accepted 2010 September 16.  Received 2010 September 15; in original form 2010 April 8}

\maketitle

\begin{abstract}
We present a study characterizing the environments of galaxies in the redshift range of $0.4 < z < 1$ based on data from the POWIR near infrared imaging and DEEP2 spectroscopic redshift surveys, down to a stellar mass of $\log~M_\ast = 10.25~M_\odot$. Galaxy environments are measured in terms of nearest neighbour densities as well as fixed aperture densities and kinematical and dynamical parameters of neighbour galaxies within a radius of 1 $h^{-1}$ Mpc. We disentangle the correlations between galaxy stellar mass, galaxy colour and galaxy environment, using not only galaxy number densities, but also other environmental characteristics such as velocity dispersion, mean harmonic radius, and crossing time.
We find that galaxy colour and the fraction of blue galaxies depends very strongly on stellar mass at $0.4 < z < 1$, while a weak additional dependence on local number densities is in place at lower redshifts ($0.4 < z < 0.7$). This environmental influence is most visible in the colours of intermediate mass galaxies ($10.5 < \log~M_\ast < 11$), whereas colours of lower and higher mass galaxies remain largely unchanged with redshift and environment. 
At a fixed stellar mass, the colour-density relation almost disappears, while the colour-stellar mass relation is present at all local densities.
We find a weak correlation between stellar mass and environment at intermediate redshifts, which contributes to the overall colour-density relation. We furthermore do not find a significant correlation between galaxy colour and virial mass, i.e., parent dark matter halo mass. Galaxy stellar mass thus appears to be the crucial defining parameter for intrinsic galaxy properties such as ongoing star formation and colour.

\end{abstract}


\section{Introduction}

The influence of environment on the properties of galaxies and their evolution has become evident over the last three decades.
Among the early evidence for an environmental effect on galaxy evolution was the discovery of the morphology-density relation \citep{Oem74,Dre80}, showing that morphologies of cluster galaxies are not randomly distributed, but depend on the distance from the cluster centre, or alternatively on the local density in which a galaxy resides. Another early study by \citet{Dav76} showed that the two-point correlation function, measured using a catalogue of preferentially selected field galaxies, depends on galaxy morphology. 
We have since learned that not only galaxy morphology depends on environment, but also the mean age of a galaxy's stellar population is affected: galaxies in low-density environments appear on average 1-2 Gyr younger than their counterparts in high density environments like galaxy clusters, or local overdensities in galaxy groups \citep{Tho05,Cle06}. This was also shown by \citet{Ber98}, who found that among a population of early-type galaxies, field ellipticals tend to present populations about 1 Gyr younger than those in clusters. Recently, \citet{Coo10} found evidence for a correlation between age and environment at fixed stellar mass, such that galaxies in higher density regions formed earlier than galaxies of similar mass in lower density environments. Similar evidence comes from studies of galaxy clustering. \citet{Ski09}
analysed the correlations between galaxy colours and the environment using marked correlation functions and argued that, in order to explain the observed colour dependence, a significant fraction of faint satellite galaxies must be blue. \citet{Ski09a} argued that much of the correlations between morphologies and the environment are explained by the colour-environment correlation.

On the other hand, the star formation histories, colours and morphologies of galaxies are very sensitive to galaxy stellar mass \citep[see e.g.][]{Bri00,Jim05,Tho05}: the higher the stellar mass of a galaxy the earlier, shorter and more intense the initial burst of star formation. The peak in the star formation activity then shifts to lower mass galaxies as cosmic time proceeds \citep[see e.g.][]{Bun06}.
The relative importance of the various factors in shaping the galaxies that we observe today is unclear. What are the roles of stellar mass, total galaxy mass, and parent dark matter halo mass? What has a stronger influence, local density or membership in a massive structure? Is being a `satellite' or a `central' galaxy an important factor, as recently suggested by e.g. \citet{vdB08} and \citet{Ski09b}?

There is considerable previous work addressing these questions. \citet{Bla05} showed that most environmental correlations can be explained by two galaxy properties, colour and luminosity, and that correlations with structural properties are less important. \citet{Cle06} find that metallicity and $\alpha$-enhancement are independent of environment, but correlate with galaxy velocity dispersion, i.e., the galaxy's dark matter halo mass. On the other hand \citet{Bla06} argue that galaxy properties are only related to the mass of its parent dark matter halo, i.e., the total mass of the structure in which a galaxy resides. \citet{Wei06} similarly emphasize the importance of group halo mass. They furthermore find that the properties of satellite galaxies are strongly correlated with those of their central galaxy.
\citet{Kau04} find that galaxy structure depends strongly on stellar mass, while star formation history is very sensitive to local density. 
Recently, \citet{Bam09} showed the presence of a morphology-density relation at fixed stellar mass using Galaxy Zoo morphologies. However, they do not find a dependence of morphology on group mass, either determined from the velocity dispersion or the integrated group light.

Looking at higher redshifts ($z\sim1$), the correlations found in the local universe seem to be either already in place or reversed. A strong dependence of galaxy colours on local projected number density was found by \citet{Coo06} up to $z\sim 1$. \citet{Cuc06} on the other hand find that the colour-density relation present at $0.25 < z < 0.60$ progressively disappears until it is undetectable at $z \sim 0.9$. The morphology-density relation was found to be already in place at $z\sim1$ over a wide range of densities from cluster cores to the field \citep{Smi05,Pos05}. 
Stellar mass seems to set an upper limit to this density dependence: \citet{Tas09} found that above M$^{\ast} \sim 10^{10.8} M_{\odot}$ morphology does not evolve or depend on environment up to $z \sim 1$, a result confirmed by \citet{Iov10}. They also argue that the lower fraction of blue galaxies in groups with respect to the field is mainly caused by luminosity selected samples biased towards blue low-mass galaxies. The colour difference largely disappears when stellar mass selected samples are used.
The relation between SFR and local density at high $z$ is controversial too. The SFR-density relation found for SDSS galaxies locally seems to be reversed at higher redshifts: \citet{Elb07} and \citet{Coo08} find galaxies with higher SFR are located in higher densities at $z \sim 1$. However, the opposite was found by \citet{Pat09}, who observe the same strong decline of SFR with density at $z \sim 0.8$ as in the SDSS.

The contradictory results of the above studies may be partly due to the problematic and manifold definitions of what a galaxy environment is, and how to quantify it. The options include (1) group or cluster membership, compared to the rather ill-defined `field', (2) local number densities, based on number counts in a fixed aperture, or the distance to the $n^{th}$ nearest neighbour, where both, the choice of aperture -size or $n$ introduce an arbitrary element, and (3) the dark matter halo mass, based on halo occupation models.

Galaxy stellar mass itself, usually regarded as an intrinsic property of a galaxy, may also depend on environment, as suggested by the stellar mass-halo mass correlation \citep[see e.g.][]{Mos10}. Massive elliptical galaxies are often found in the cores of galaxy clusters, or at high local densities, while lower mass spirals are located in the outskirts of larger structures or in small groups, as our Local Group. However, this traditional view is not necessarily always the case: massive ellipticals are also found in the field \citep[e.g.][]{Col01}, and low-mass galaxies with elliptical morphology are found preferentially in high local densities \citep[e.g.][]{Rob07}. 
It is hence not clear if, and how, a galaxy's stellar mass depends on the environment and how this dependence evolves with redshift.

High redshift studies have additional problems: deep surveys usually cover a small area, suffering from the problem of cosmic variance. Another issue is the limited availability of spectroscopic redshifts. Photometric redshifts, even quite accurate ones, make it difficult to accurately define environments since they smear out structures along the line of sight. This is especially problematic at low densities \citep{Coo05}.

We approach this problem with the combination of a deep infrared survey covering a large field of view, the POWIR survey, and the DEEP2 redshift survey providing the spectroscopic redshifts for a large number of objects. Additionally, our galaxy sample is selected in the $K$-band.
The near infrared bands are a much better proxy for stellar mass than optical bands, since they are sensitive to the light of old, low-mass stars and less sensitive to the effects of dust. Optical colours correspond to rest-frame UV light, which traces young and hot stellar populations, and thus is more sensitive to star formation than stellar mass. The near-IR selection therefore corresponds to a selection based on stellar mass rather than star-formation. This allows us to study correlations with stellar mass without the colour bias of optical luminosity selected samples \citep{Con08}.

Section~\ref{the sample} describes the two surveys and the resulting samples used in this study. The ways in which we characterise a galaxy's environment is explained in Section~\ref{character}. Section~\ref{results} presents our results, Section~\ref{discussion} discusses its implications and Section~\ref{conclusion} summarizes our most important findings.
We assume the standard $\Lambda$CDM cosmology and a flat universe with $\Omega_\Lambda = 0.73$,  $\Omega_M = 0.27$ and a relative Hubble constant $h = H_0 / 100$ is used.

\section{The sample}\label{the sample}

\subsection{The POWIR and DEEP2 surveys}
This study is based on data obtained by the Palomar Observatory Wide-Field Infrared (POWIR) survey described in \citet{Con08}. The goal of this survey is to construct a $K$-band selected and stellar mass limited sample covering a large field of view, and containing a large number of galaxies with  spectroscopic redshifts.
The POWIR survey covers four fields: the Extended Groth Strip \citep[EGS,][]{Dav07} and three additional fields subsequently referred to as Field 2, 3 and 4. All four fields have been observed by the DEEP2 (Deep Extragalactic Evolutionary Probe) redshift survey \citep{Dav03}. 

Deep $J$ and $K_s$-band imaging was obtained with the WIRC camera on the Palomar 5m telescope. The survey consists of 75 pointings covering an area of $\sim$1.5 deg$^2$ with an average 5$\sigma$ depth reaching down to $K_{s,vega} = 20.2$. For a detailed description of the observations and data reduction procedures see \citet{Bun06}.
The corresponding optical data come from the 3.6m Canada-France-Hawaii Telescope (CFHT) using the CFH12K camera. Imaging in the bands $B$, $R$ and $I$ was taken, reaching a 5$\sigma$ depth of $R_{AB} = 25.1$. The details of these observations and their reduction are given in \citet{Coi04}.
The seeing is comparable in the infrared and optical data ($FWHM \sim 1$ arcsec). All photometry was measured within a 2 arcsec diameter aperture.

Spectroscopic redshifts were obtained with the Keck DEIMOS spectrograph \citep{Fab03} as a part of the DEEP2 redshift survey \citep{Dav03}. Targets were selected down to $R_{AB} < 24.1$ independent of their colour in the EGS field, whereas in fields 2, 3 and 4 targets were additionally selected based on $(B-R)$ and $(R-I)$ colour, focusing on selecting galaxies at $z > 0.7$.  The spectroscopic sampling rate is about 60\% in all four fields. Together with the spectroscopic success rate of $\sim 70 \%$ this yields an overall spectroscopic completeness of $\sim 40 \%$ in all four fields over a redshift range of $0.4 < z < 1$. The different target selection criteria in fields 2, 3, 4 and the EGS lead to an overall smaller number of galaxies in the redshift interval $0.4 < z < 0.7$, since objects in this range come mainly from the EGS field. This redshift interval thus has a smaller surveyed volume relative to the higher redshift range of $0.7 < z < 1$.

The photometric catalogue of galaxies detected in the $K_s$-band images with the SExtractor package \citep{Ber96} was then matched to the DEEP2 redshift catalogue. The matching between the $R$-band selected redshift catalogue and the K-band selected photometric catalogue was done within a 1 arcsec radius. This results in a very low rate of spurious matches of 1-2\% due to the low surface densities in both catalogues \citep{Bun06,Con08}.
The spectroscopic sample is magnitude limited at $R_{AB} = 24.1$ due to the DEEP2 target selection limit, whereas the $K_s$-band selected photometric catalogue has a magnitude limit of $R_{AB} = 25.1$. This leads to a lower percentage of galaxies with spectroscopic redshifts ($\sim 22\%$, see \citet{Con07}) in the photometric catalogue than the spectroscopic completeness of $40\%$ quoted above. For all galaxies in the $K_s$-band selected photometric sample, which have no measured spectroscopic redshift, we measure photometric redshifts. This procedure is described in the following section. For a more detailed description of the sample selection and an extensive discussion of the completeness we refer the reader to \citet{Con07}.

\subsection{Additional photometric redshifts}

The spectroscopic redshift sample is complemented by photometric redshifts measured from the optical and IR photometry described above.  For galaxies that meet the spectroscopic target selection, but were not observed due to instrumental constraints, the neural network code ANNZ \citep{Col04} was used. This includes all galaxies brighter than $R_{AB} = 24.1$, which were either not sampled spectroscopically ($\sim 40\%$ of galaxies in our catalogue with $R_{AB} = 24.1$) or which have no reliable spectroscopic redshift measurement ($\sim 30\%$ of spectroscopically observed galaxies). This sample of galaxies without spectroscopic redshifts ($\sim60\%$ of the total $K_s$-band selected sample) cover the same magnitude range as galaxies with spectroscopic redshift, which can therefore be used at a training set for the ANNZ neural network code.
Photometric redshifts of galaxies fainter than $R_{AB} = 24.1$ were measured with the BPZ package \citep{Ben00}, which uses a Bayesian approach, including information about the likelihood of a certain redshift-brightness combination, taken from the distribution of galaxies in the Hubble Deep Field \citep{Ben00}. BPZ is well suited for determining photometric redshifts of faint galaxies, for which no comparison sample with spectroscopic redshifts is available.

The photometric redshift accuracy is good out to $z\sim1.4$ with an uncertainty of $\Delta z /(1+z) = 0.07$ \citet{Con07}. For a detailed description of the photometric redshift estimates we refer the reader to \citet{Con07,Con08} and \citet{Bun06}.
The numbers of galaxies in the full sample (spectroscopic redshifts plus additional photometric redshifts) and the spec-z sample (secure spectroscopic redshifts only) are as follows: the full sample has 50,119 galaxies, while the spec-z sample comprises 10,682 galaxies.

\subsection{Stellar mass measurements}

Stellar masses were estimated by fitting the optical and near infrared photometric datapoints with synthetic spectral energy distributions (SEDs) from \citet{Bru03} spanning a variety of galaxy ages, metallicities and star formation histories. We use an exponentially declining
star formation history with an e-folding time between 0.01 and 10 Gyr and ages of the star formation onset between 0 and 10 Gyr. The metallicities span a range of 0.0001 to 0.05.
The typical uncertainties of the model SED fitting is $\sim$ 0.1-0.2 dex. Adding to this the photometric errors yields a total stellar mass uncertainty of 0.2-0.3 dex. We note that for intermediate age stellar populations the stellar masses can be systematically overestimated. However, as shown in \citet{Con07}, for the present sample the masses are lower by on average 0.07 dex only, when using models with AGB-TP stars. Another source of systematic error is the choice of IMF. Here, the IMF of \citet{Cha03} is used. The details of the stellar mass measurements are fully described in \citet{Bun06}.

\citet{Bun06} and \citet{Con07} also estimate the completeness limits in stellar mass from the $K_s$-band detection limit at different redshifts by placing model galaxies at the high redshift end of each redshift bin. For a redshift of $z\sim1$ they find that the full sample (including galaxies with photometric redshifts) is 100\% complete down to a stellar mass of $\log~M_\ast=10.25$. Galaxies down to this stellar mass and redshift limit are detected independently of their colour. Our sample is therefore not strongly biased towards bluer galaxies at high $z$. We use a stellar mass limit of $\log~M_\ast=10.25$ and a redshift limit of $z=1$ throughout this paper. 

Out of the 50,119 galaxies in our whole catalogue 14,563 are above the stellar mass limit $\log~M_\ast > 10.25$ and below the redshift limit $z\leq1$ and hence are included in this study. 4101 of these galaxies have spectroscopic redshifts. The final full sample therefore has 14,563 galaxies, and the spec-z sample has 4101 galaxies.

%---------------------------begin Figure 1------------------------------------------
\begin{figure*}
\includegraphics[width=0.32\textwidth]{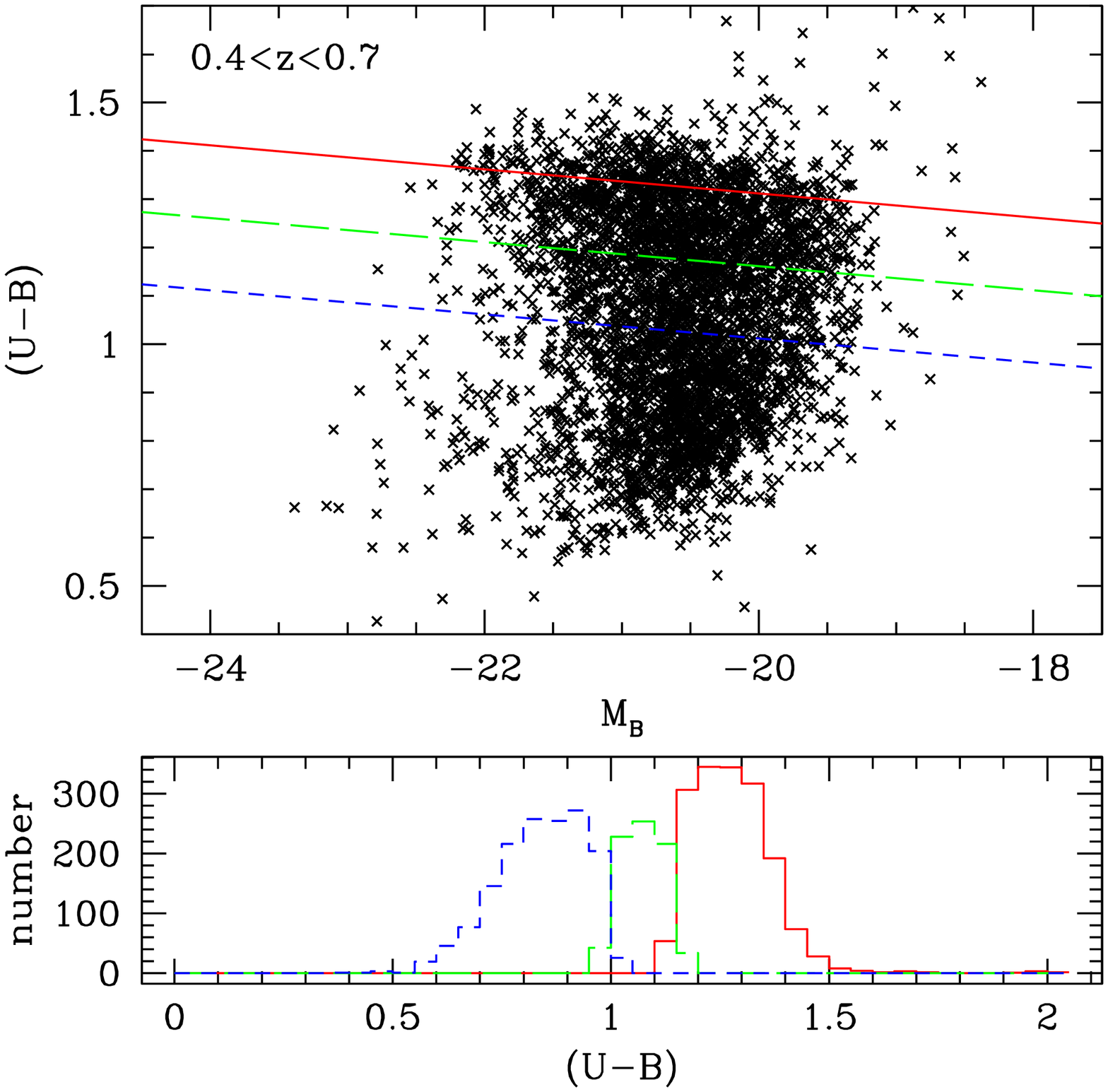}
\includegraphics[width=0.32\textwidth]{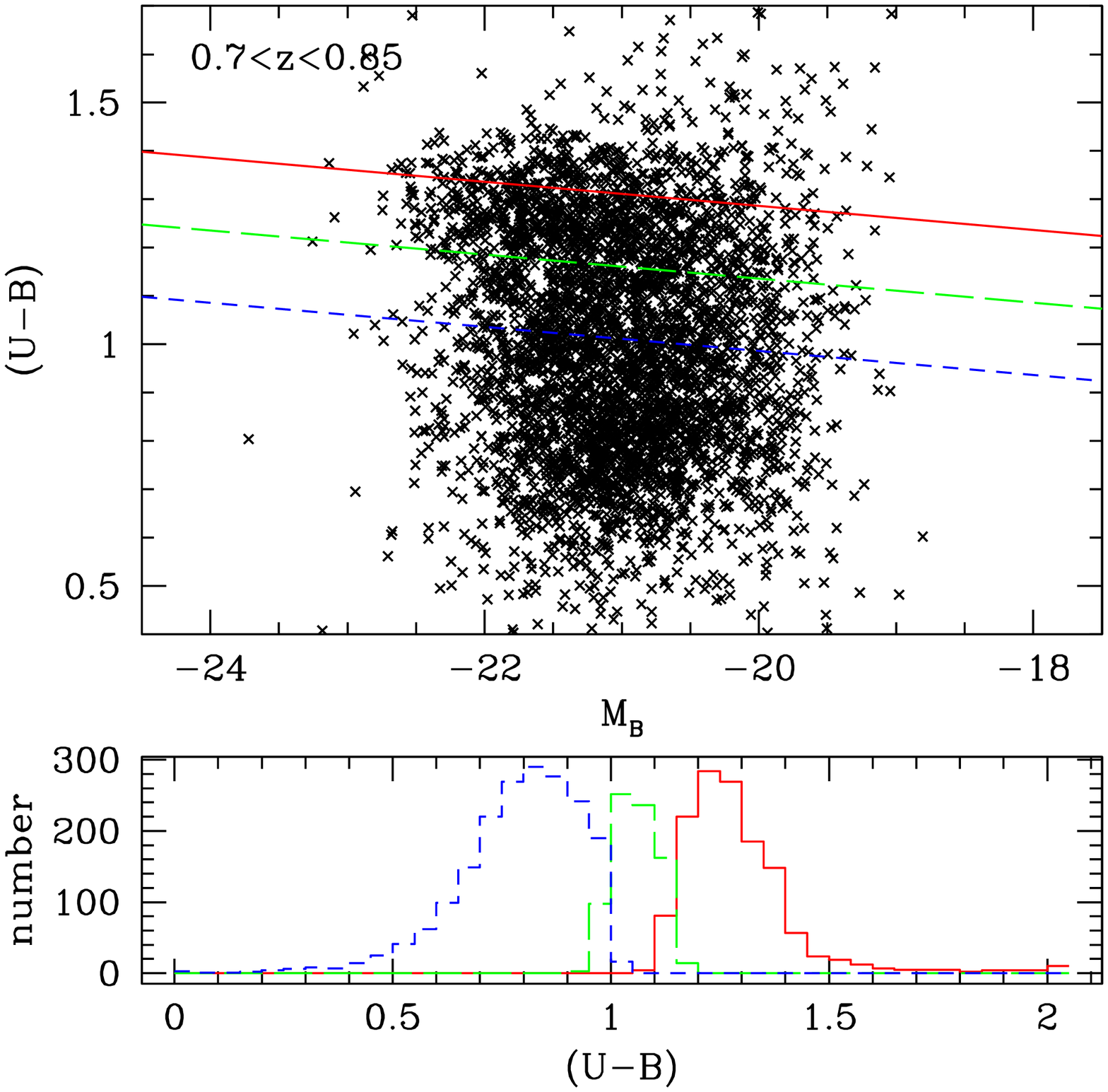}
\includegraphics[width=0.32\textwidth]{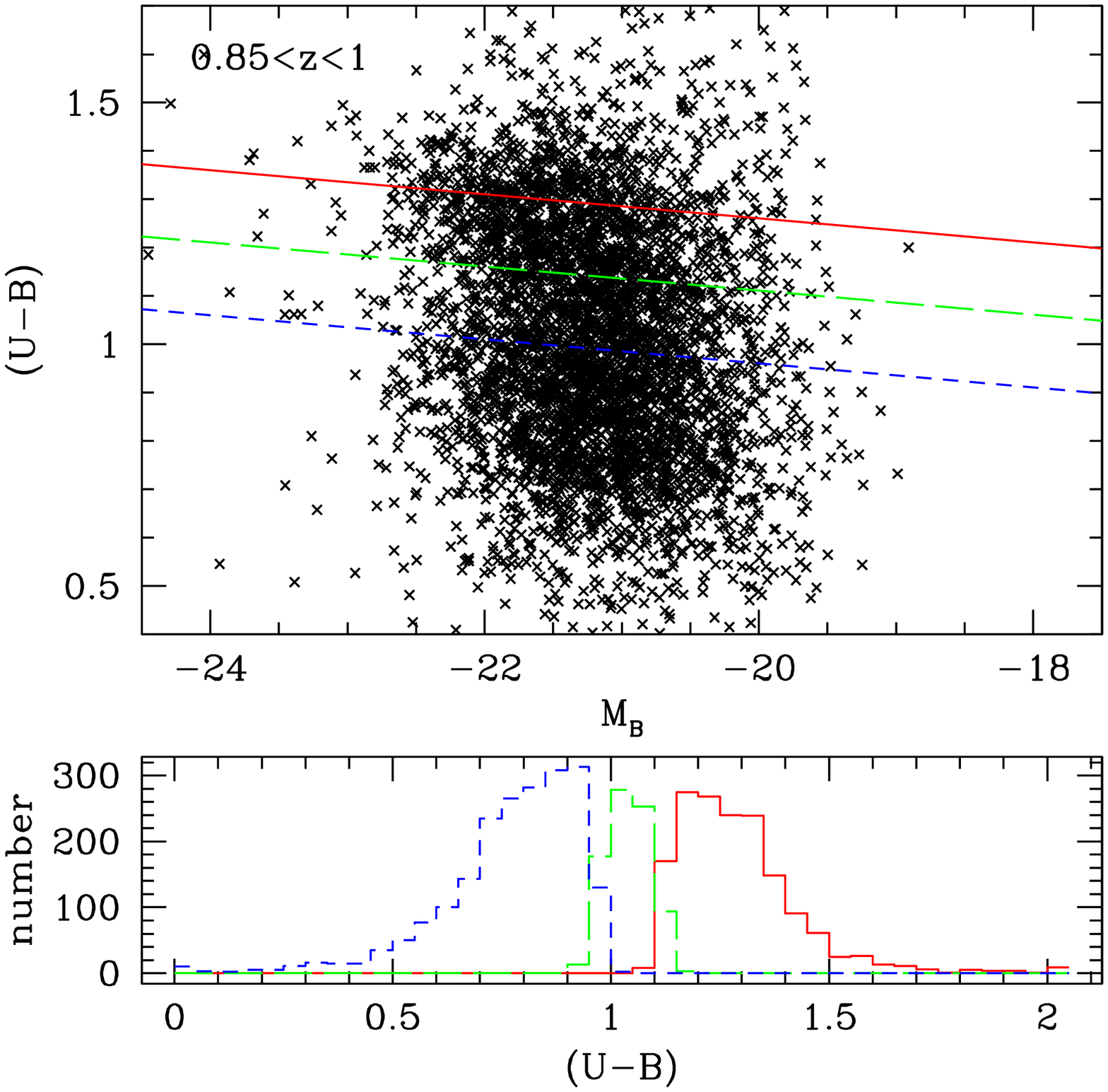}
\caption{Colour magnitude diagram (CMD) of the full sample (14,563 galaxies) in three redshift bins. The red (solid) line indicates the location of the red sequence found by \citet{Wil06}, while the green (dotted) and blue (dashed) lines are the limits below (bluer than) which galaxies are considered green and blue respectively. The small panels below each CMD shows the colour distribution of blue, green, and red galaxies.
\label{fig1}}
\end{figure*}
%---------------------------end Figure 1------------------------------------------

\subsection{Rest-frame $(U-B)$ colours and colour fractions $f_{blue}$,$f_{green}$ and $f_{red}$}\label{fblue}

Rest-frame $(U-B)$ colours were derived by \citet{Wil06} in the Vega magnitude system. All magnitudes and colours used in this paper are converted into the AB magnitude system. The transformation between the Vega and AB system is a linear shift given by the AB magnitude of Vega in the respective filter. The $(U-B)$ colours used in this study can be transformed in the following way: $(U-B)_{AB}$ = $(U-B)_{Vega}$ + 0.83 \citep{Wil06}.

The blue fraction is defined as the fraction of galaxies that have not reached the red sequence yet. It is computed with the magnitude dependent colour limit defined by \citet{Wil06}. By fitting the red sequence of the colour-magnitude relation of the DEEP2 sample they found a clear distinction between the red sequence and the blue cloud up to about $z \sim 1$. The colour limit separating these two populations is defined to be $\sim0.25$ magnitudes blueward of the red sequence, given by the following equation (shifted into the AB magnitude system):

\begin{equation}
(U-B) = -0.032~(M_B+21.52)+1.284
\end{equation}

\noindent Here we not only compute fractions of galaxies on the red sequence, $f_{red}$, and in the blue cloud, $f_{blue}$, but we also quantify the number of galaxies located in the transition region between blue and red, i.e., the fraction of galaxies in the `green valley', $f_{green}$. To identify these green galaxies we assume a width of the green valley of $\Delta (U-B) = 0.15$ starting at 0.15 magnitudes blueward of the red sequence. Red galaxies are then defined as

\begin{equation}
(U-B) > -0.032~(M_B+21.52)+1.284-0.15
\end{equation}

\noindent and blue galaxies as

\begin{equation}
(U-B) < -0.032~(M_B+21.52)+1.284-0.3
\end{equation}

\noindent while all galaxies in between are green.

Figure~\ref{fig1} shows the colour limits in the colour magnitude diagram in three redshift bins. The red line corresponds to the red sequence of \citet{Wil06}, while the green line shows the limit blueward of which galaxies are considered to lie in the green valley. Galaxies below the blue line are considered blue cloud systems.

\section{Characterisation of galaxy environments}\label{character}

This section describes the different environment measures we use in this study. We use two different approaches: the first one is based on measuring the distance to the $n$-th nearest neighbours of each galaxy within a redshift slice, while the second one uses all galaxies within a certain physical radius and radial velocity interval.

\subsection{Nearest neighbour densities} 

\citet{Coo07} measured the third nearest neighbour densities of galaxies in the DEEP2 sample, which we utilize here. The choice of $n$ for $n$-th nearest neighbour distances was found to have a weak influence on the resulting density values \citep{Coo05}. \citet{Coo05} also investigated possible effects of the selection function of the DEEP2 survey and conclude that it does not introduce any environmental bias, at least up to $z \sim 1$.

The method used to determine the nearest neighbour densities is fully described in \citet{Coo06} and \citet{Coo07}. Only galaxies with spectroscopic redshifts are used in measuring local densities. First the projected distance $D_3$ to the third nearest neighbour within a radial velocity window of $\pm1000$ km s$^{-1}$ is measured. This is then converted into a surface density $\Sigma_3 = 3/(\pi D_3^2)$. To account for different sampling rates of the DEEP2 survey at different redshifts, $\Sigma_3$ is divided by the mean surface density $\langle \Sigma_3 \rangle$ in slices of $\Delta z = 0.04$. This yields the relative overdensity $(1+\delta_3)$, which is not affected by redshift-dependent incompleteness. Note that the ratio $\Sigma_3/\langle\Sigma_3\rangle$ is not denoted $\delta_3$, but by $(1+\delta_3)$. $\delta_3$ itself is the overdensity defined as $(\Sigma_3 - \langle\Sigma_3\rangle)/\langle\Sigma_3\rangle$. Using $\log~(1+\delta_3)$ is convenient for separating the sample into over- and underdense regions: if $\log~(1+\delta_3)$ is positive, the galaxy is located in an overdense region, while a negative $\log~(1+\delta_3)$ corresponds to an underdense area relative to the mean density at each redshift.

To minimize edge effects, we exclude galaxies closer than 1 $h^{-1}$ Mpc to a survey border. Out of the 4101 galaxies in the spectroscopic redshift sample, 486 galaxies are within the 1 $h^{-1}$ Mpc border region and excluded from the further analysis. The sample used in the local density analysis then consists of 3615 galaxies. 

%spec-z: with density: 3917, more than 1Mpc from border: 3146
%full: with density: 4549, more than 1Mpc from border: 3615

\subsection{Environment within a fixed aperture}

\begin{figure}
\includegraphics[width=0.477\textwidth]{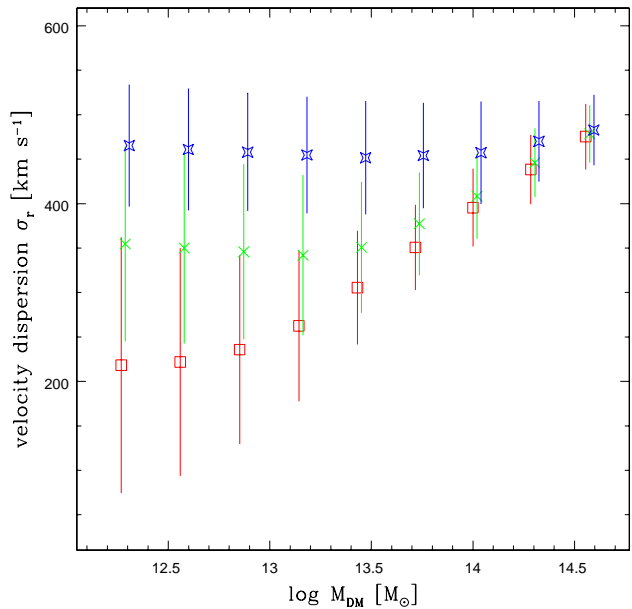}
\caption{Comparison between dark matter halo mass $M_{\rm DM}$ and velocity dispersion $\sigma_r$ computed within different circular apertures and a radial velocity interval of $\Delta v = \pm 1000$ km~s$^{-1}$. The results for the different apertures are shown as blue stars (5 $h^{-1}$ Mpc), green crosses (2 $h^{-1}$ Mpc) and red squares (1 $h^{-1}$ Mpc). The errorbars give the $1\sigma$ dispersion in each bin. A small shift in $M_{\rm DM}$ is applied for a better visualization of the errorbars.
\label{fig2}}
\end{figure}

In addition to the nearest neighbour densities, which are expressed as relative overdensities, the absolute local density for each galaxy is calculated by counting neighbouring galaxies within a certain fixed radius (or aperture), and a fixed radial velocity interval. This search radius is chosen to be 1 $h^{-1}$ Mpc, which gives the best correlation (as we will show below) between number of neighbours, their velocity dispersion and the parent dark matter halo mass based on the Millennium Simulation \citep{Spr05}. 
We show this by using a light-cone catalogue produced from the full simulation within a box of 500 $h^{-1}$ Mpc on each side. The dark matter haloes are populated with galaxies according to the halo occupation distribution (HOD) models of \citet{Ski09}. Galaxy luminosities are computed according to the model described in \citet{Ski06}. The light-cone catalogue is fully described in \citet{Ski09}. It comprises positions, distances and magnitudes for 954,212 simulated galaxies down to an absolute magnitude limit of $M_{r}-5\log h = -19$. The distances include the peculiar  motions of the galaxies and are converted into radial velocities using the relative Hubble constant $h = 1$, as it is used in the simulation. The number densities and galaxy velocity dispersion within different fixed apertures of 1, 2, and 5 $h^{-1}$ Mpc and a velocity window of $\pm$1000 km~s$^{-1}$ is computed. 
Figure~\ref{fig2} shows the relation between the measured velocity dispersion of all galaxies within the fixed aperture and velocity interval as a tracer of the halo mass and the halo mass from the simulation itself. The different apertures are shown as blue stars (5 $h^{-1}$ Mpc), green crosses (2 $h^{-1}$ Mpc) and red squares (1 $h^{-1}$ Mpc). The errorbars show the $1~\sigma$ dispersion in each bin.
Within an aperture of 5 $h^{-1}$ Mpc there is no relation between halo mass ($M_{\rm DM}$) and velocity dispersion ($\sigma_r$): $\sigma_r$ has a value of around 500 km s$^{-1}$ at all halo masses. At 2 $h^{-1}$ Mpc $\sigma_r$ starts to be sensitive to the halo mass, but the relation flattens off at $\log~M_{\rm DM}\sim13.5 M_\odot$ and $\sigma_r \sim 350$ km s$^{-1}$. Using 1 $h^{-1}$ Mpc the relation becomes steeper and we are able to trace the parent dark matter halo mass down to $\sim 10^{12.5} M_\odot$ or $\sigma_r \sim 200$ km s$^{-1}$ (Figure~\ref{fig2}). 

The radial velocity interval within which neighbouring galaxies are included is set to $\Delta v = \pm 1000$ km~s$^{-1}$. This choice may sound arbitrary, however it is motivated by the distribution of galaxies in our survey: the vast majority of galaxies in the DEEP2 survey are either isolated or located in galaxy groups, not in clusters \citep{Ger07}. A velocity of 1000 km~s$^{-1}$ corresponds to the escape velocity of a massive galaxy group with a velocity dispersion of $ \sigma_r \sim 300$ km~s$^{-1}$.

%---------------------------begin Figure 2------------------------------------------
%-----------------------------end Figure 2 ----------------------------------------

From all galaxies found within the fixed aperture and velocity interval we then compute the following quantities:

\begin{enumerate}
\renewcommand{\theenumi}{\arabic{enumi}.}

\item Radial velocity dispersion $\sigma_r$ and the virial mass $M_{\rm vir}$ \indent \indent 
as a halo mass estimator; 

\item Harmonic radius $R_H$ as a compactness estimator; 

\item Crossing time $t_c$ as a dynamical state estimator; 

\item The offset in projection and velocity, relative to the \indent \indent 
average, $R$;

\item  Number of neighbours $N_{\rm 1Mpc}$ within 1 $h^{-1}$ Mpc as a \indent \indent 
richness estimator. 

\end{enumerate}

\noindent
The definitions of the above quantities are given in the following. The radial velocity dispersion of a group structure is defined as:

\begin{equation}
\sigma_r = \left[\frac{\sum_{i}^{N_{\rm 1Mpc}} (v_i - v_{\rm mean})^2}{{N_{\rm 1Mpc} - 1}}\right]^\frac{1}{2} \\
\end{equation}

\noindent where $v_i$ is the radial velocity of each galaxy and $v_{\rm mean}$ is the mean velocity of all $N_{\rm 1Mpc}$ galaxies within the 1 $h^{-1}$ Mpc radius. 

The mean harmonic radius ($R_H$) is a measure of the compactness. It gives the mean projected separation between all objects in a galaxy's environment, i.e., the compactness of the environment. This is related to the local density, such that a galaxy in a high local density will have a small $R_H$. However, also a low local density can result in a small $R_H$, if the galaxy is part of an isolated close pair or triple. $R_H$ is calculated from the projected separations $R_{ij}$ between the i-th and j-th galaxy:

\begin{equation}
R_H =  \left[\frac{\sum_{i} \sum_{j<i} 1 / R_{ij}}{(N_{\rm 1Mpc} (N_{\rm 1Mpc}-1))/2}\right]^{-1}
\end{equation}

\noindent From the velocity dispersion ($\sigma_r$) and the mean harmonic radius ($R_H$), the crossing time ($t_c$) is computed following the definition of \citet{Rood78}:

\begin{equation}
t_c = \frac{2 R_H}{\sqrt{3} \sigma_r}
\end{equation}

\noindent The crossing time describes how long it would take the average galaxy moving with the velocity $\sigma_r$ to cross its local environment defined by $R_H$.  Multiplying $t_c$ with H$_0$ gives the crossing time in units of the age of the universe, independent of the choice of H$_0$. The inverse of this quantity then estimates how often the galaxy could have crossed its environment within the universe's lifetime. The crossing time is also an indicator of dynamical equilibrium: crossing times of about $\frac{1}{5}$ of a Hubble time indicate that a structure is possibly virialized \citep{Fer90}. \\

As an indicator of a galaxy's parent dark matter halo mass, the virial mass ($M_{\rm vir}$) was computed following \citet{Heisler85}:

\begin{equation}
M_{\rm vir} = \frac{3 \pi \sigma_r^2 R_H}{2 G}
\end{equation}

\noindent where G is the gravitational constant. $M_{\rm vir}$ is a straightforward and useful halo mass estimator, given that the structure is in dynamical equilibrium. However, one has to keep in mind that calculated with a small number of objects the virial mass can be underestimated by a factor of 3-5 \citep{Heisler85}.

The $R$-parameter has been proposed by \citet{ZM00} in order to facilitate the comparison between the distribution of galaxies in projection and in velocity space. It indicates the position of a galaxy relative to the average offset from the projected centre of space and velocity.
It is defined as 
\begin{equation}
R = \sqrt{\left(\frac{d}{\delta_d} \right)^2 + \left(\frac{|v_{\rm pec}|}{\delta_{|v_{\rm pec}|}}\right)^2}
\end{equation}

\noindent where $\delta_d$ and $\delta_{|v_{\rm pec}|}$ denotes the rms deviations in projected distance and peculiar velocity respectively of all galaxies within the 1 $h^{-1}$ Mpc radius. A galaxy with a large distance $d$ from the geometric centre or a large offset from the mean velocity $v_{\rm pec}$ will yield a large value of $R$ while an average galaxy should have $R \sim \sqrt{2}$. A small $R$ then indicates that the galaxy is closer to the centre in space and velocity than the average galaxy, which can be interpreted as being more likely at rest in (or close to) the centre of the potential well of the parent structure or dark matter halo.

\section{Results}\label{results}

In the following we present our results in two ways, comparing the full sample ($z_{spec}$ and $z_{phot}$ combined) and for galaxies with secure spectroscopic redshifts only. The photometric redshifts are not suited to compute dynamical quantities, since the typical uncertainties of $\Delta z / (1+z) \sim 0.03$ are much larger than the typical velocity dispersion of galaxy groups. However, for comparison reasons we plot the obtained values next to the ones obtained by using spectroscopic redshifts only in Figures~\ref{fig4}~and~\ref{fig5}.
As described in Section~\ref{the sample} our sample consists of galaxies down to a stellar mass of $M_\ast = 10.25$ up to a redshift of $z=1$.
To avoid edge effects, galaxies with a distance from the border of the surveyed field of less than 1 $h^{-1}$ Mpc were excluded in the following analysis.
First we compare the two local density estimators, $N_{\rm 1Mpc}$ and $(1+\delta_3)$ as described in Section~\ref{character}, then we describe the evolution of the environmental characteristics ($\sigma_r$ and $M_{\rm vir}$, $R_H$, $t_c$, $R$ and $N_{\rm 1Mpc}$) and the rest-frame colours with cosmic time. We then investigate the correlation between colours and fraction of red, green and blue galaxies with local density, halo mass and stellar mass. And finally, to disentangle the influence of stellar mass and local density on galaxy colour, the sample is split in different bins of stellar mass and local density.

\subsection{Richness $N_{\rm 1Mpc}$ versus relative overdensity $(1+\delta_3)$}

As described in Section~\ref{character}, two different methods are used to estimate richness/density: the number of neighbours within a fixed aperture, $N_{\rm 1Mpc}$, and the nearest neighbour density, $(1+\delta_3)$, based on the distance to the $3^{rd}$ nearest neighbour, measured by \citet{Coo07}.
The same population of galaxies is used to measure the two local density estimators. Only galaxies with secure spectroscopic redshifts are used when comparing  $(1+\delta_3)$ with $N_{\rm 1Mpc}$.
Although the two values should roughly agree overall, they trace different things: $N_{\rm 1Mpc}$ is equivalent to an absolute co-moving density, while $(1+\delta_3)$ is a relative {\it over-}density, i.e., density normalized by the mean density in the galaxy's respective redshift interval ($\pm0.04$). 
The nearest neighbour density is an adaptive measurement that is well suited to characterize local galaxy concentrations on small scales, which might not be identified as density enhancements by the fixed aperture density. Furthermore it gives a continuous range of densities, while the fixed aperture density can by construction only have discreet values, increasing the intrinsic uncertainty of the measurement. On the other hand the $n$-th nearest neighbour density is more sensitive to interlopers, since due to the variable area, the resulting density can change significantly due to the presence of a single object, especially when a small $n$ is used. 

We decided not to convert $N_{\rm 1Mpc}$ into units of Mpc$^{-2}$ but to leave it in number of galaxies within the aperture. Since the aperture is the same for all galaxies, this has no effect on the results, but allows us to easily determine how many galaxies our results are based upon. 

%---------------------------begin Figure 3------------------------------------------
\begin{figure}
\includegraphics[width=0.35\textwidth, angle=270]{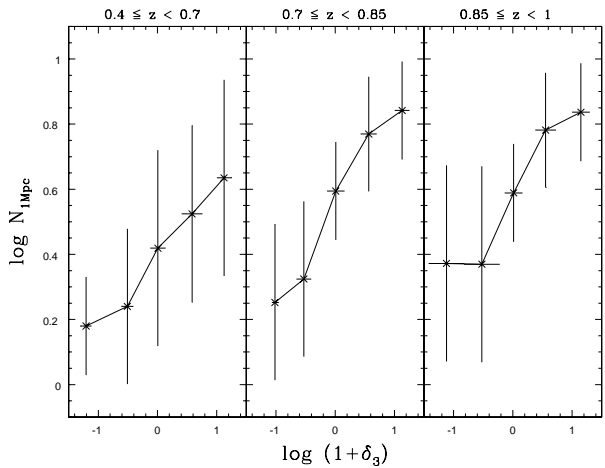}
\caption{Comparison between relative over-density $\log~(1+\delta_3)$ and richness $\log~N_{\rm 1Mpc}$. The errorbars show the 1$\sigma$ spread in each bin.
\label{fig3}}
\end{figure}
%-----------------------------end Figure 3 ----------------------------------------

Figure~\ref{fig3} compares $\log~N_{\rm 1Mpc}$ with $\log~(1+\delta_3)$ in the three redshift ranges. The two measures agree on average. 
To quantify the correlation between the two densities we compute Spearman rank correlation coefficients $\rho$. The value of $\rho$ ranges between $-1 \leq \rho \leq 1$, where $\rho=1$ ($\rho=-1$) means that two variables are perfectly correlated (anti-correlated) by a monotonic function. Completely uncorrelated variables result in $\rho=0$. Taking into account the sample size in each bin we estimate the significance of the value of $\rho$ using the conversion from correlation coefficient to z-score \citep{Fie57}.
The correlation coefficients and their significance in the three redshift bins are $\rho_{0.4-0.7}=0.31$ at $\sim3\sigma$, $\rho_{0.7-0.85}=0.50$ at $\sim8\sigma$, and $\rho_{0.85-1}=0.44$ at $\sim7\sigma$. However, in the most over- and underdense environments the linear correlation becomes flatter. This shows that the nearest neighbour density $(1+\delta_3)$ is better suited for tracing the high and low ends of the relative density distribution.

% no_mass_binning.sm
% n(0.4-0.7) & $\langle M_\ast \rangle$ & $\langle z \rangle$ & n(0.7-0.85) & $\langle M_\ast \rangle$ & $\langle z \rangle $ & n(0.85-1) & $\langle M_\ast \rangle$ & $\langle z \rangle$ \\
%1841     10.43      0.5589    1628   10.43      0.7789   1504  10.44      0.9251
%1459     10.80      0.5615    1465   10.80      0.7786   1687  10.81      0.9240
%606      11.15      0.5741    805   11.16      0.7821    1074  11.15      0.9294
%76       11.48      0.5761    137   11.49      0.7759    169   11.49      0.9279

%plot_groups_zNEW.sm
\begin{table*}
\begin{scriptsize}
\caption{Numbers of galaxies, mean stellar mass and mean redshift in each bin of stellar mass and redshift in Figure~\ref{fig4} and Figure~\ref{fig5} for the spec-z only sample (a) and the full sample (b).}
\label{tab1}
\begin{flushleft} 
\begin{tabular}{l c ccc c ccc c ccc c ccc}
\hline
(a) & & n(M1) & $\langle \log~M_\ast \rangle$ & $\langle z \rangle$ & & n(M2) & $\langle \log~M_\ast \rangle$ & $\langle z \rangle $ & & n(M3) & $\langle \log~M_\ast \rangle$ & $\langle z \rangle$ & & n(M4) & $\langle \log~M_\ast \rangle$ & $\langle z \rangle$ \\
z1~~ & & 97   &    10.38  &    0.5802 & & 226  &  10.75   &   0.6032 & & 116  &  11.18   &   0.6124 & & 11  &   11.63  &   0.6062 \\
z2~~ & & 346   &    10.37  &    0.7846 & & 740  &  10.76   &   0.7833 & & 370  &  11.18   &   0.7864 & & 24  &   11.60  &   0.7840 \\
z3~~ & & 239   &    10.37  &    0.9264 & & 521  &  10.75   &   0.9202 & & 323  &  11.17   &   0.9245 & & 16  &   11.62  &   0.9132 \\
\hline
(b)  & & n(M1) & $\langle \log~M_\ast \rangle$ & $\langle z \rangle$ & & n(M2) & $\langle \log~M_\ast \rangle$ & $\langle z \rangle $ & & n(M3) & $\langle \log~M_\ast \rangle$ & $\langle z \rangle$ & & n(M4) & $\langle \log~M_\ast \rangle$ & $\langle z \rangle$ \\
z1~~ & & 1174   &   10.37   &   0.5589 & & 1914  &  10.73  &   0.5664  & & 602  &  11.18   &   0.5797 & & 28  &  11.63   &   0.5815 \\
z2~~ & & 1070   &   10.38   &   0.7802 & & 1887  &  10.74  &    0.7799  & & 853  &  11.18   &   0.7827 & & 54  &  11.61   &   0.7794 \\
z3~~ & & 990   &   10.37   &   0.9248 & & 2164  &  10.75  &    0.9241  & & 1168  &  11.18   &   0.9294 & & 72  &  11.65   &   0.9175 \\
\hline
\end{tabular}

stellar mass bins: M1: $\log~M_\ast < 10.5$, M2: $10.5 < \log~M_\ast < 11$, M3: $11 < \log~M_\ast < 11.5$, M4: $\log~M_\ast > 11.5$ \\
redshift bins: z1: $0.4 \leq z <0.7$, z2: $0.7 \leq z < 0.85$, z3: $0.85 \leq z < 1$ \\
\end{flushleft} 
\end{scriptsize}
\end{table*}

\subsection{Galaxy environment correlated with stellar mass and cosmic time}

%---------------------------begin Figure 4------------------------------------------
\begin{figure*}
\includegraphics[width=0.485\textwidth]{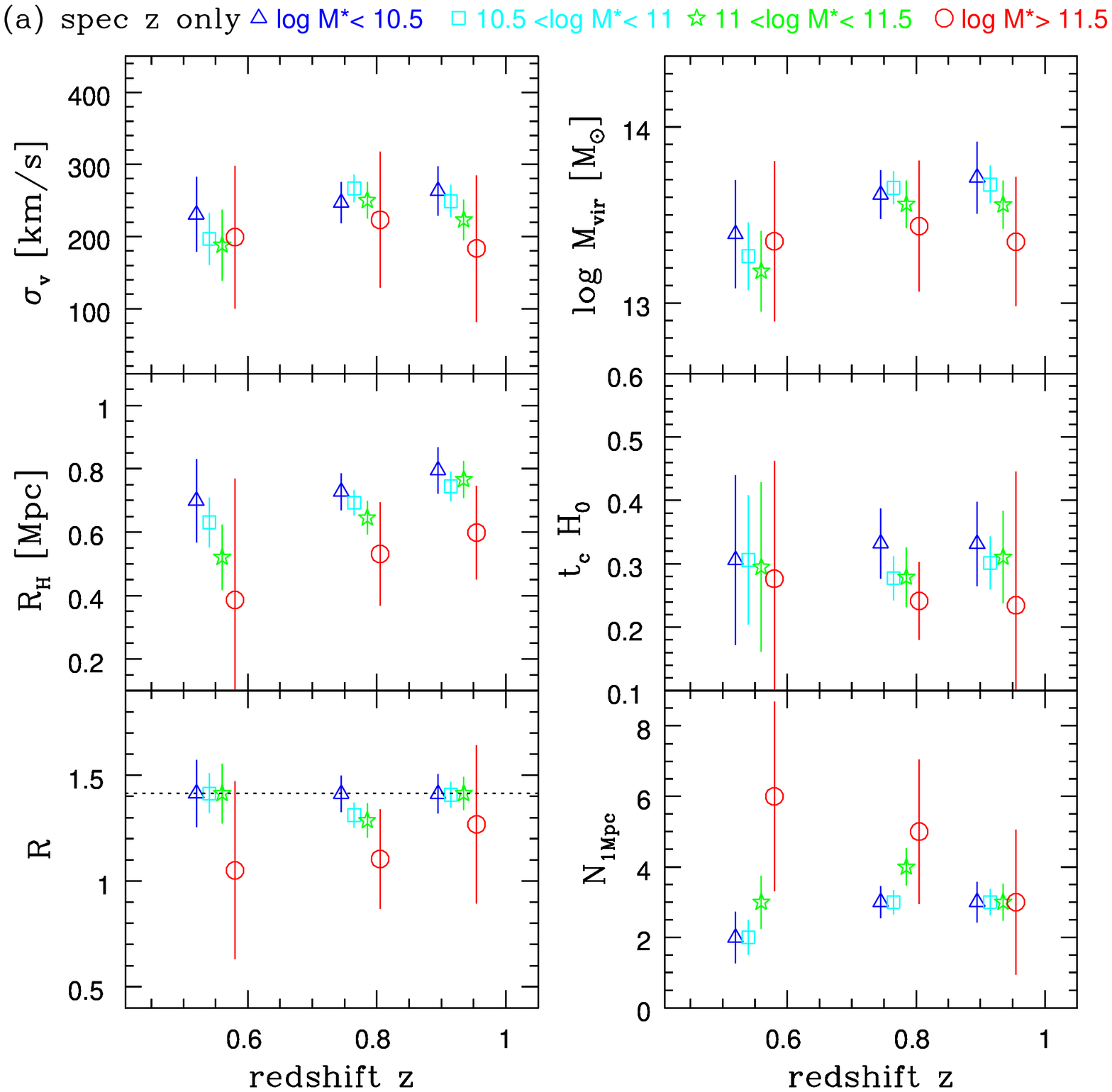}
\includegraphics[width=0.485\textwidth]{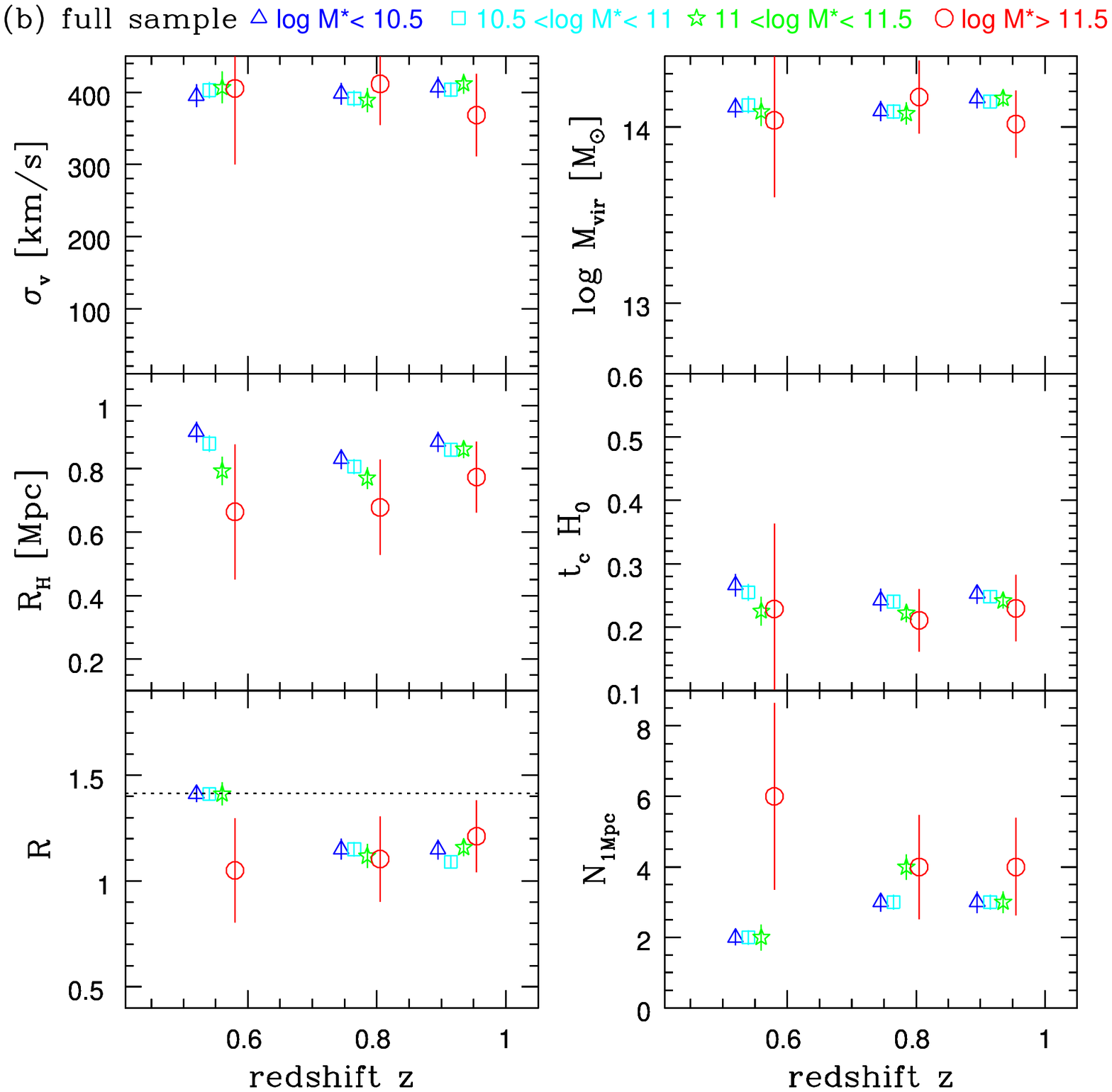}
\caption{Environmental properties of galaxies at different stellar masses as a function of redshift. Left side: spectroscopic redshifts only; right side: full sample. Note that velocity related quantities measured with the full sample (including galaxies with photometric redshifts) are not reliable measurements and we show them only for comparison reasons. Top left: radial velocity dispersion $\sigma_r$; top right: virial mass $\log~M_{\rm vir}$; mid left: harmonic radius $R_H$; mid right: crossing time in units of the Hubble time $t_c H_0$; bottom left: $R$ parameter, the dotted line shows the average value of $R\sim \sqrt{2}$ (see text); bottom right: number of galaxies $N_{\rm 1Mpc}$. The sample is sliced into four stellar mass and three redshift bins. The stellar mass bins are colour-coded: red circles ($\log~M_\ast > 11.5$), green stars ($11.5 \geq \log~M_\ast > 11$), cyan boxes ($11 \geq \log~M_\ast > 10.5$), and blue triangles ($\log~M_\ast < 10.5$). The redshift bins are $0.4 \leq z < 0.7$, $0.7 \leq z < 0.85$, and $0.85 \leq z < 1$. The datapoints in each bin are offset in redshift direction for clarity. The errorbars represent the $3\sigma$ error of the mean in each bin.\label{fig4}}
\end{figure*}
%-----------------------------end Figure 4 ----------------------------------------

The different environmental characteristics, as described in Section~\ref{character}, are plotted in Figure~\ref{fig4} as a function of $z$. 
Note that all the quantities are measured with galaxies within 1 $h^{-1}$ Mpc radius and within a radial velocity interval of $\Delta v = \pm 1000$ km s$^{-1}$ as described above.
To investigate the evolution of these quantities with redshift, as well as possible differences in the evolution of galaxies of different stellar mass, the sample is split in four stellar mass bins and three redshift bins. These three redshift bins (used throughout the paper) are: $0.4 \leq z< 0.7$, $0.7 \leq z< 0.85$ and $0.85 \leq z< 1$. The stellar mass bins are indicated at the top of the figure with the respective colour: $\log~M_\ast < 10.5$ in blue,  $10.5 < \log~M_\ast < 11$ in cyan, $11 < \log~M_\ast < 11.5$ in green and $\log~M_\ast > 11.5$ in red.  Table~\ref{tab1} gives the numbers of galaxies as well as the mean values of stellar mass and redshift in each bin of stellar mass and redshift.
The sum of the number of galaxies over all bins is lower than the sample size quoted in Section~\ref{the sample}, due to galaxies that have no neighbours within the search radius and therefore no measured environmental parameters. These are about 25\% of galaxies in the spectroscopic redshift sample and about 20\% of galaxies in the full sample.

%---------------------------begin Figure 5------------------------------------------
\begin{figure*}
\includegraphics[angle=270,width=0.485\textwidth]{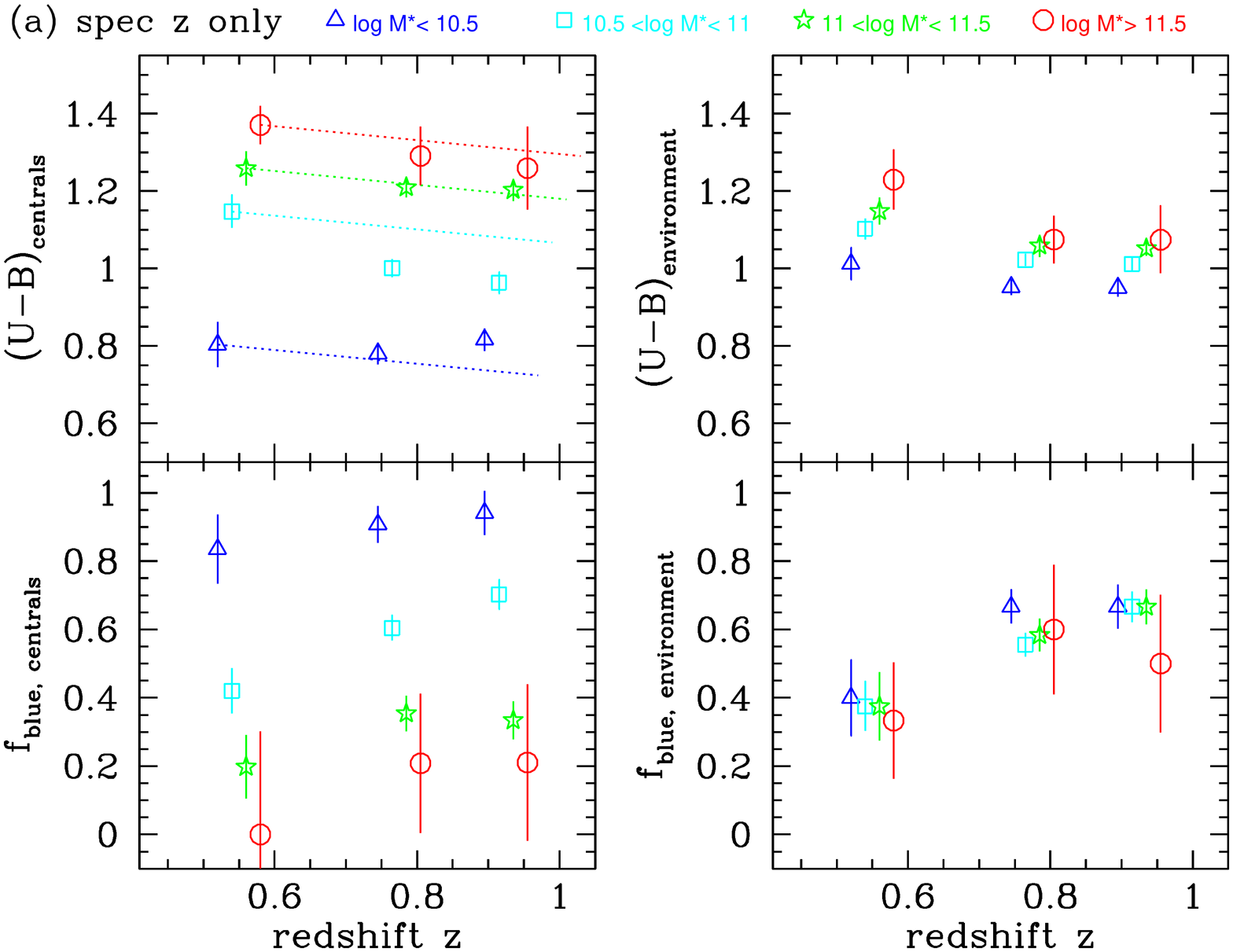}
\includegraphics[angle=270,width=0.485\textwidth]{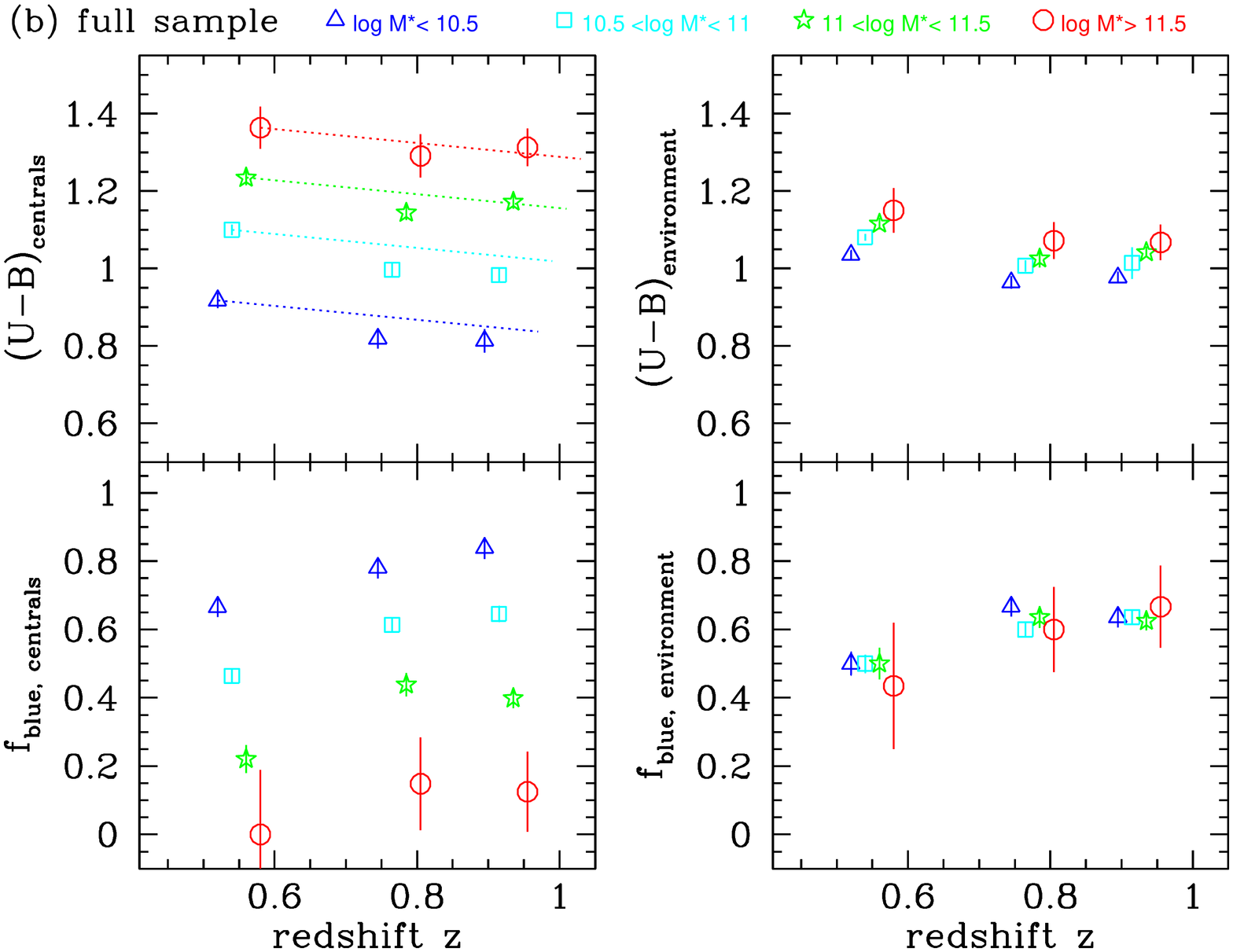}
\caption{Colours of galaxies with different stellar masses and their environments as a function of redshift. Left side: spectroscopic redshifts only; right side: full sample. Top left: $(U-B)$ colour of `central' galaxies; top right: $(U-B)$ colour of galaxies in the central's environment; bottom left: blue fraction $f_{blue}$ of `central' galaxies; bottom right: blue fraction of galaxies in the central's environment. The sample is sliced in four stellar mass and three redshift bins, as in Fig.~\ref{fig4}. The dashed lines trace the expected colour evolution of a passively evolving stellar population formed at high $z$ (see text). The errorbars represent the $3\sigma$ error of the mean in each bin. \label{fig5}}
\end{figure*}
%-----------------------------end Figure 5 ----------------------------------------

\subsubsection{The parent dark matter halo: $\sigma_r$ and $M_{\rm vir}$}\label{sigma}

Figure~\ref{fig4}, top row, shows the median $\sigma_r$ and $M_{\rm vir}$ in three redshift bins.
The range of velocity dispersions $\sigma_r$  is found to be largely similar at all redshifts. There are no significant changes with redshift or stellar mass. The same is seen for the virial mass $M_{\rm vir}$. We do not see a significant trend for more massive galaxies to be situated in more massive structures as measured by $\sigma_r$ at $0.4 < z < 1$, neither do we find obvious indications for a growth of halo mass over time. However, this might be due to the large uncertainties of $M_{\rm vir}$ due to the generally small numbers of galaxies found ($\langle N \rangle < 10$ at all redshifts and masses).
The comparison between the spectroscopic sample (left side) and the full sample (spectroscopic plus photometric redshifts, right side) illustrates that photometric redshifts are not suited for this kind of study. The velocity dispersion calculated including galaxies with photometric redshifts is systematically overestimated, with $\sigma_r \sim 400$ km s$^{-1}$, as expected for a random distribution of velocities within a $\pm1000$ km s$^{-1}$ window. The same can be seen for $M_{\rm vir}$ which is of course related to $\sigma_r$.

\subsubsection{Compactness and dynamical state: mean harmonic radius $R_H$ and crossing time $t_c$}\label{rh_tc}

Figure~\ref{fig4}, mid row, shows the median $R_H$ and $t_c$ values in three redshift bins.
We find that $R_H$ does vary slightly with redshift as well as with galaxy stellar mass. There is a trend that more massive galaxies reside in more compact environments, i.e., at lower $R_H$.
The typical crossing times $t_c$ have large scatters and seem to be largely independent of redshift and galaxy stellar mass. The median $t_c$ are of the order of 0.2-0.3 Hubble times, which is consistent with the time-scale in which relaxation can take place \citep{Fer90}.
There is a trend for more massive galaxies to be located in more relaxed or more dynamically evolved structures, i.e., in structures with a shorter crossing time, however, this trend is only significant in one redshift bin ($0.7 \leq z < 0.85$).  
$R_H$ is less affected by the use of photometric redshifts than $t_c$, since it is not velocity related, while $t_c$ is measured from the velocity dispersion. $R_H$ is only affected by the inclusion of foreground/background galaxies due to the use of photo-z's, while $t_c$ is directly affected by the artificially high velocity dispersion in the full sample.

\subsubsection{Distribution of galaxies in phase space: the $R$ parameter}\label{Rparam}

The bottom left panel of Figure~\ref{fig4} shows the evolution of the $R$ parameter with redshift.
The most massive galaxies show on average consistently smaller $R$ values than the lower mass galaxies. They seem to occupy a special position in phase space, although the uncertainties are large due to the small number of galaxies with stellar masses above $10^{11.5}~M_\odot$. However,
the low $R$ suggests that the most massive galaxies sit closer to the centre of their potential well or dark matter halo than an average galaxy in our sample. 
This result is consistent with what was found for groups dominated by bright ellipticals in the local universe: \citet{ZM00} find that the brightest group galaxies (BGGs) are more likely to lie close to the centre of projected space and velocity than the rest of the group population.  
We find possible redshift evolution in the value of $R$. Between $0.8 < z \leq 1$ galaxies with different stellar masses are indistinguishable in their position in phase space, while at lower redshifts ($z \leq 0.7$) the most massive galaxies with $M_\ast \geq 10^{11.5}~M_\odot$ have $R \sim 1$ compared to $R \sim 1.4$ for lower mass galaxies. This difference is significant at about $3\sigma$.
Including galaxies with photometric redshifts dilutes the result: now all galaxies are more `central' than galaxies in their environment, but simply because they were selected to be so, and more of the included galaxies around them are random foreground or background objects. 

\subsubsection{Richness of the structure: number of neighbours $N_{\rm 1Mpc}$}\label{richness}

The average number of neighbours found for each galaxy within our 1 $h^{-1}$ Mpc and $|\Delta v| \leq$ 1000 km~s$^{-1}$ limits are shown in the bottom right panel of Figure~\ref{fig4}. The numbers are similar for galaxies with lower stellar masses ($\log~M_\ast < 11$), while the more massive galaxies have generally more neighbours. This trend is most significant (3$\sigma$ confidence level) for the most massive galaxies, but is also visible for galaxies in the range $11 < \log~M_\ast < 11.5$. Furthermore, the most massive galaxies show increasing numbers of neighbours with decreasing redshift, while for the rest of the galaxies the median number of neighbours is roughly constant with redshift. The same trend can be seen for the relative overdensity $\log~(1+\delta_3)$ which will be discussed later in this paper.

\subsection{Rest-frame colours and the blue fraction of galaxies as a function of cosmic time}\label{colour-z}

In the following section we show how the average $(U-B)$ colour and blue fractions change with cosmic time. Figure~\ref{fig5} shows the evolution of galaxy rest-frame colours and blue fraction with redshift, divided in the same four stellar mass bins as used before.
The colours do not evolve strongly with redshift, as already noticed by \citet{Wil06}. 
The dotted lines show the expected evolution of $(U-B)$ with redshift for an old passively evolving stellar population taken from \citet{vDF00}. Each line is adjusted to the low redshift data point of each stellar mass bin, tracing back the expected colour of passively evolving galaxies at higher redshifts. The expected change is very small and the average colours indeed do not change very much with redshift.
However, there is a slight colour evolution, especially for intermediate mass galaxies ($10.5 < \log~M\ast < 11$). Their mean colour changes from $(U-B)\sim0.95$ at $z \sim 0.9$ to $(U-B)\sim1.15$ at $z \sim 0.6$. 
As can be seen in Figure~\ref{fig5} there is a very strong dependence of $(U-B)$ colour on stellar mass, which we will investigate in more detail in Section~\ref{colour-mass}. The same is true for the blue fraction which strongly depends on galaxy stellar mass. The blue fraction $f_{blue}$ increases only slightly with redshift for low and high-mass galaxies. For intermediate mass galaxies, however, $f_{blue}$ evolves strongly from 70\% at $z \sim 0.9$ to 45\% at $z \sim 0.6$ (see Figure~\ref{fig5}), analogue to the evolution in $(U-B)$. 

The mean $(U-B)$ colour and blue fraction of galaxies found in each galaxy's environment are shown in Figure~\ref{fig5}. They are denoted by $(U-B)_{environment}$ and $f_{blue,environment}$, in contrast to $(U-B)_{centrals}$ and $f_{blue,centrals}$ for the average colour and blue fraction of galaxies in each redshift bin. Note that in this context `centrals' does not imply that these galaxies are the central galaxy of their parent dark matter halo, but is used as distinction to the colour and blue fraction of a galaxy's environment.
The environment-colour and blue fraction shows if and how `satellite' galaxies react to the properties ($M_\ast$, local density) of the central galaxy in their halo. The top right panel of Figure~\ref{fig5} shows that the mean colour of the environment also depends slightly on the `central' galaxy's stellar mass. Higher mass galaxies are surrounded by redder neighbours. However, this dependence is much weaker than the primary dependence of a galaxy's colour on its own stellar mass.
Since the most massive galaxies are more likely central galaxies, we would expect to see a different behaviour of the environment-colour and environment-blue fraction in the highest $M_\ast$ bin.
Examining Figure~\ref{fig5} it appears this could be the case: environment-colour (and blue fraction) seems to rise (and drop) faster with redshift than the other three lower stellar mass bins. The significance of this difference is about 3$\sigma$.

The inclusion of galaxies with photometric redshift does not lead to large changes in the results. Rest-frame colour is not as sensitive to photometric redshift errors as the environmental characteristics discussed above.

\subsection{Rest-frame colours and colour fractions correlated with environment}

In this section we investigate the dependence of the $(U-B)$ colour and fractions of blue, red and green galaxies on environment, i.e., local density and halo mass. As described above, the nearest neighbour density $(1+\delta_3)$ is a better tracer of the extremes in the density distribution and is therefore used in the following analysis, instead of the aperture density $N_{\rm 1Mpc}$. The colour-density relation is well studied up to $z\sim1$ \citep[see e.g.][]{Coo06,Cuc06,Iov10} and can be seen as an analogue to the morphology-density relation \citep[see e.g.][]{Dre80}. To mimic the separation into different morphological types, we plot the fraction of red, blue and green galaxies, analogous to ellipticals, spirals and S0 galaxies. Colour does not perfectly correspond to morphology, but since we lack the morphological information we take colour as a proxy of morphology.

We also compare the colour and red, blue and green fractions with the virial mass $M_{\rm vir}$. The virial mass is computed from the velocity dispersion, which is a good indicator of a galaxy's parent dark matter halo mass (see Section~\ref{character} and Figure~\ref{fig2}). 
The dark matter halo can influence the properties of its galaxies through various processes like ram pressure stripping, harassment, galaxy merging or strangulation. The likelihood of these processes which either enhance or suppress star formation depends on the mass of the halo, where the processes that quench star formation like ram pressure stripping, which is proportional to the density of the intra cluster medium and the square of the velocity dispersion \citep{GG72}, or `harassment' (repeated high velocity encounters, see e.g. \citet{Moo96}) are occurring in massive dark matter haloes. 
Galaxy merging on the other hand which can trigger star formation preferentially occurs in smaller groups or lower mass haloes, since cold gas and low relative velocities are required to efficiently trigger star formation \citep[see e.g.][]{Mih94}.
We would then expect a possible connection between galaxy colour and the mass of its parent dark matter halo, such that higher mass dark matter haloes are populated by on average redder galaxies. On the other hand, the process of strangulation \citep{Lar80}, which involves the stripping of hot gas only, could occur in low-mass and high-mass haloes. \citet{Ski09b} argue that the near independence of the satellite colour distribution on halo mass may be evidence of strangulation, since it requires a process that occurs independently of halo mass. In this context it is interesting to test if colour and colour fractions show a dependence on virial mass.

%---------------------------begin Figure 6------------------------------------------
\begin{figure*}
\includegraphics[width=0.485\textwidth]{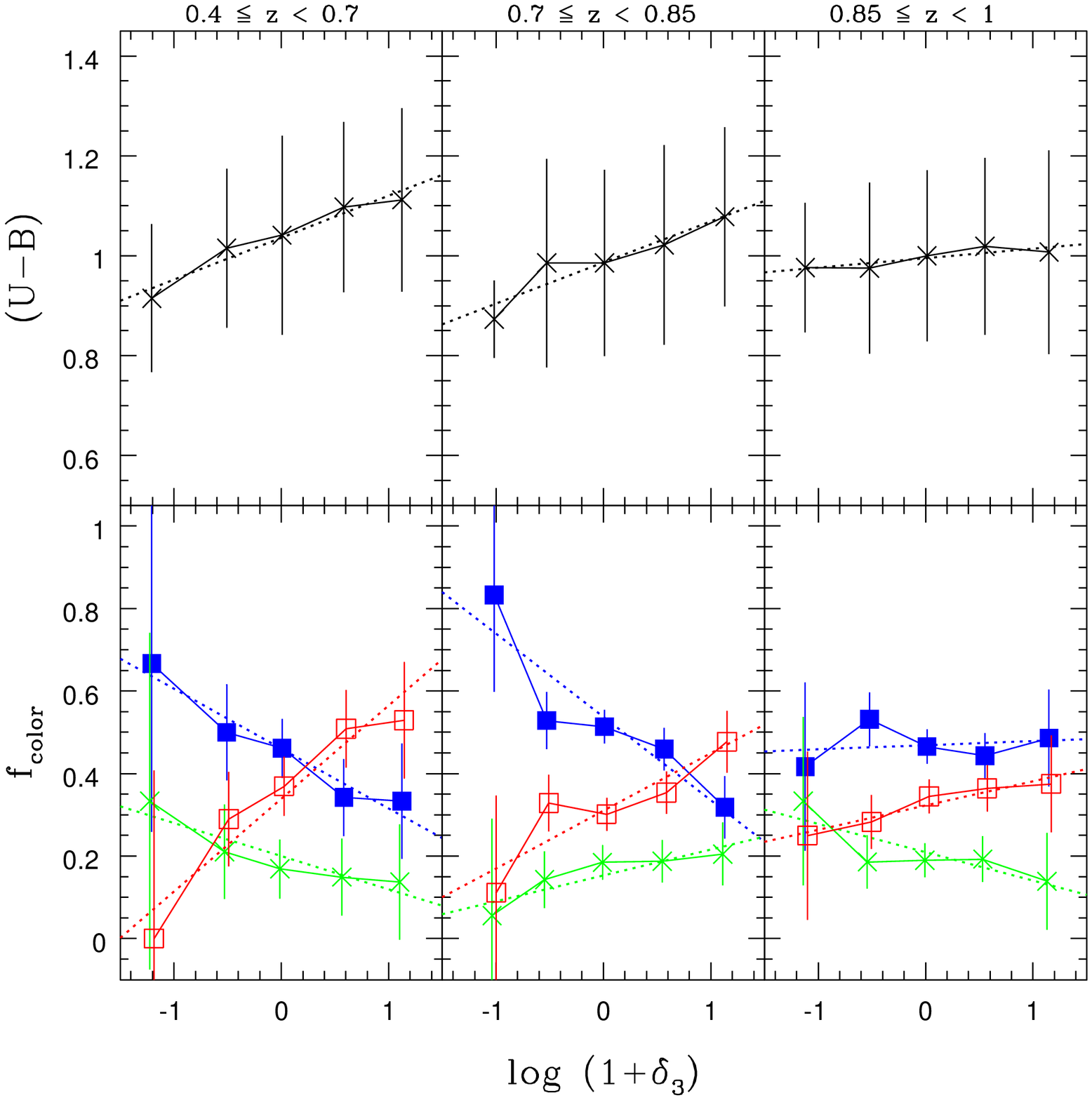}
\includegraphics[width=0.485\textwidth]{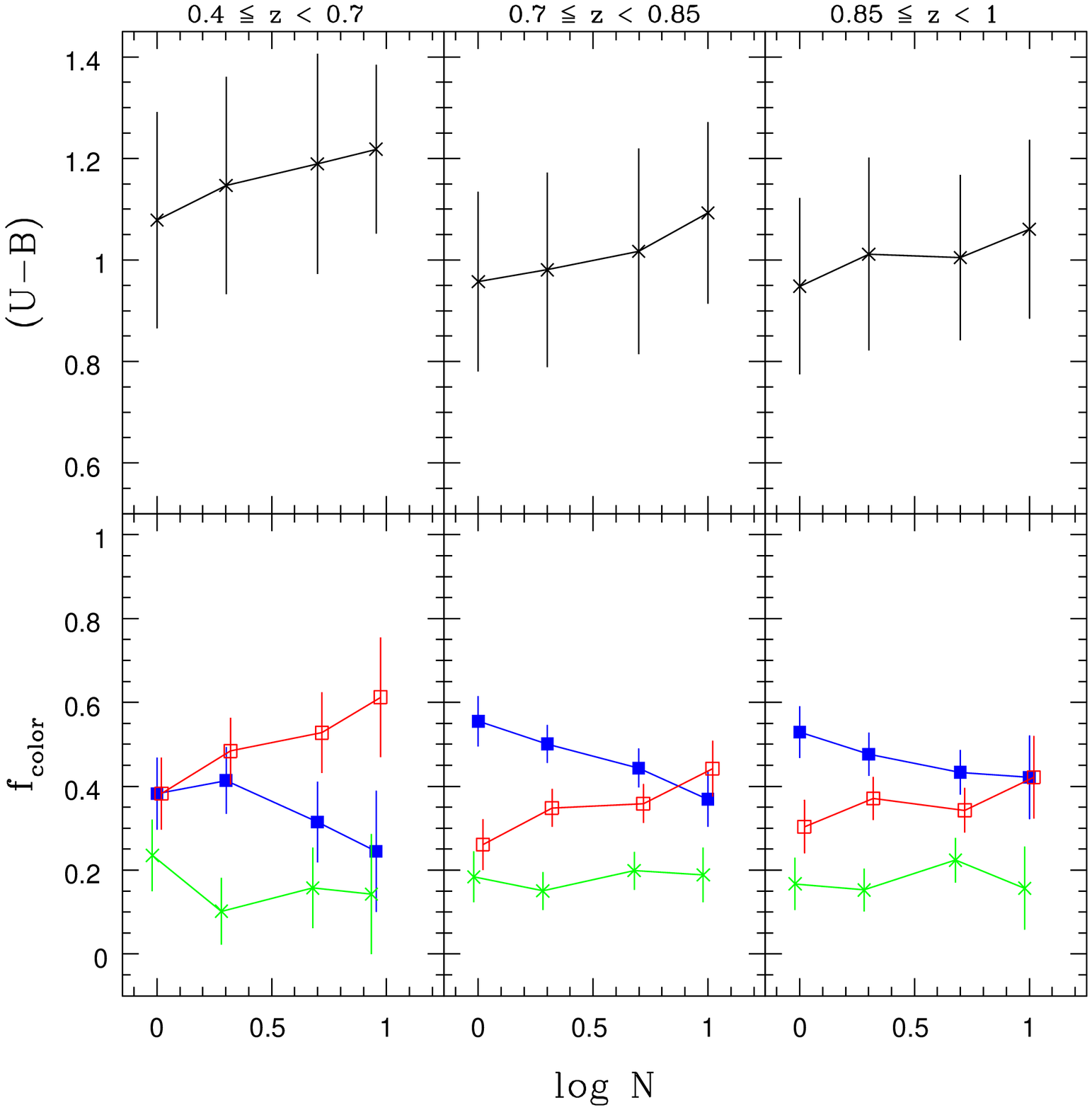}
\includegraphics[width=0.485\textwidth]{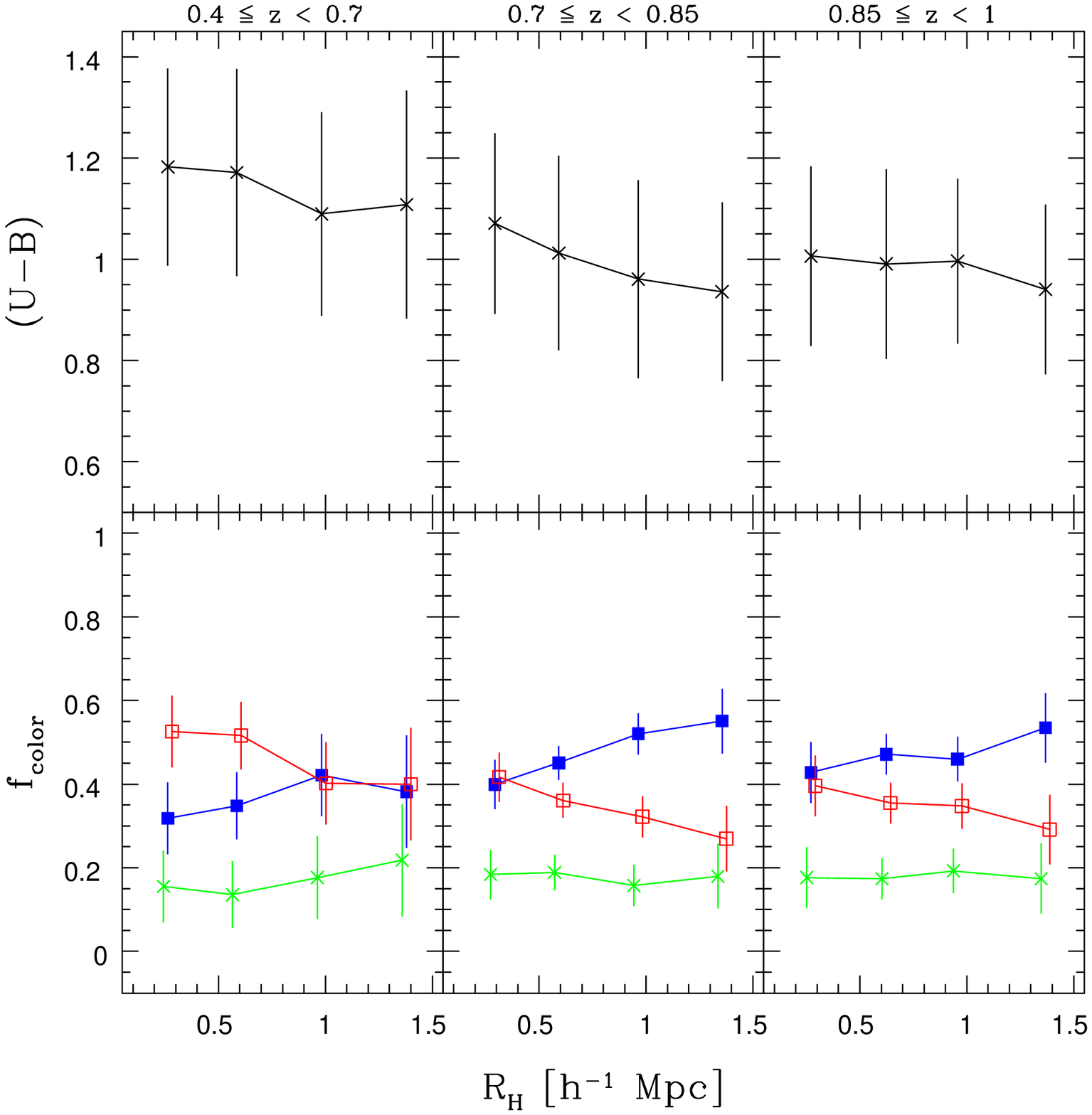}
\includegraphics[width=0.485\textwidth]{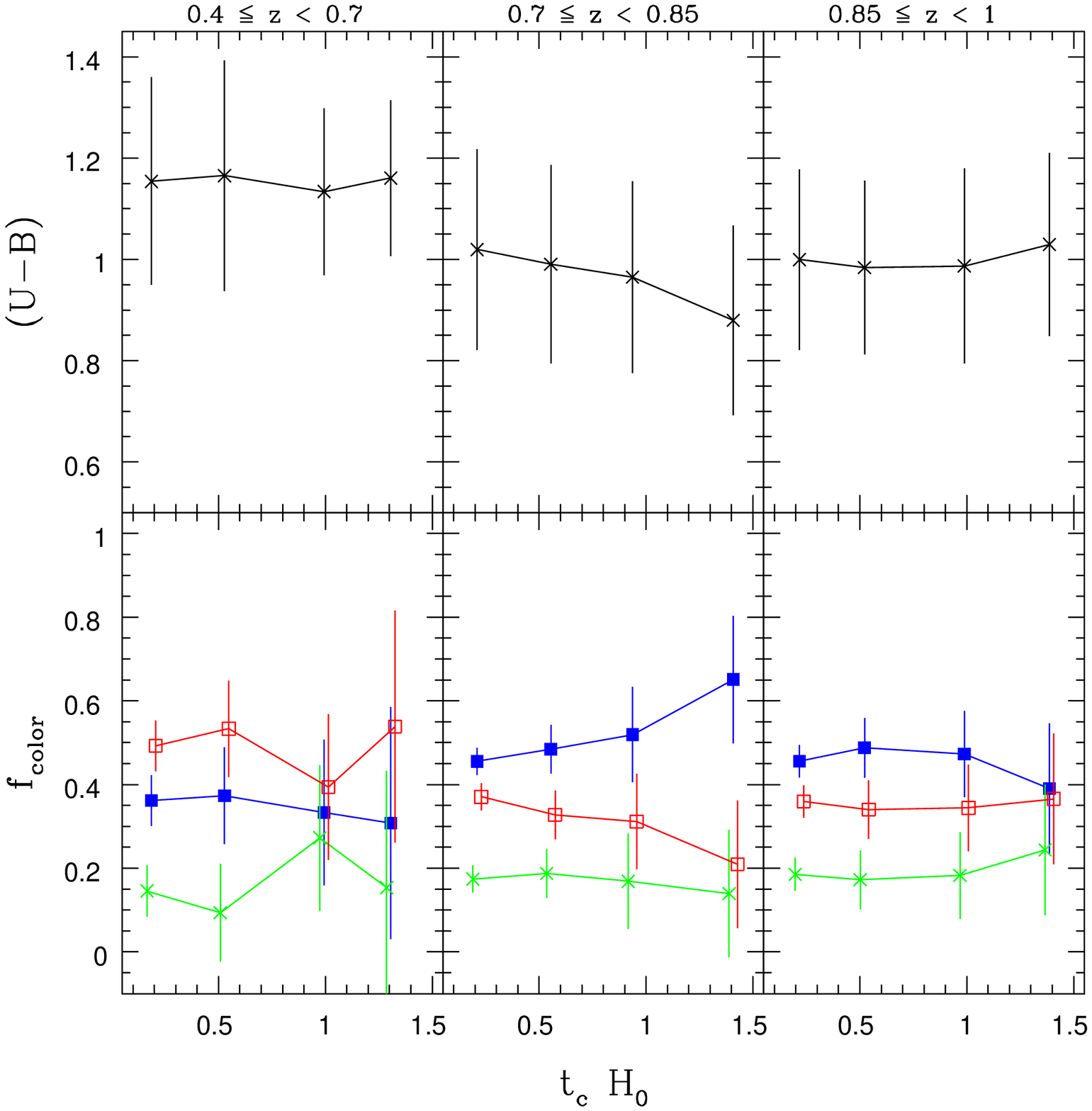}
\caption{$(U-B)$ colour and the fraction of blue, red and green galaxies as a function of different environmental characteristics in three redshift bins (as in Fig.~\ref{fig4}). Top left: nearest neighbour density $\log~(1+\delta_3)$; top right: number of neighbours $\log~N_{\rm 1Mpc}$; bottom left: mean harmonic radius $R_H$; bottom right: dimensionless crossing time $t_c~H_0$. Each plot is divided in mean $(U-B)$ (top panels) and colour fractions (bottom panels), $f_{blue}$ (blue solid boxes), $f_{red}$ (red open symbols) and $f_{green}$ (green crosses). See text for the definition of colour fractions. The errorbars represent the $1\sigma$ spread in each bin. The dashed lines are least-square fits to the datapoints.
\label{fig6}}
\end{figure*}
%-----------------------------end Figure 6 ----------------------------------------

\subsubsection{$(U-B)$ and $f_{colour}$ as a function of local density and compactness}

First, we investigate the colour-density relation for our entire sample with no stellar mass binning, analogous to other studies of the colour-density relation \citep{Cuc06,Coo07,Cas07} and the morphology-density relation \citep{Smi05,Pos05}.
Mean $(U-B)$ colours and the fractions of red, green and blue galaxies are plotted as a function of local density $\log~(1+\delta_3)$ in Figure~\ref{fig6}, top left. There is a dependence of both colour and colour fractions on the local density. The blue fraction $f_{blue}$ decreases from $\sim$0.65 in underdense environments to $\sim$0.35 at high overdensity. There is also a related increase of $f_{red}$ with local density, while the fraction of green galaxies $f_{green}$ is constant, within the errors. The colour-density relation also evolves with redshift, steepening significantly with later cosmic time. The decrease of $f_{blue}$ with $\log~(1+\delta_3)$ is significant at $\sim 2\sigma$ in the two lower redshift bins and disappears at $z>0.85$. We compute the Spearman rank coefficient $\rho$ to quantify the correlation between $(U-B)$ and $\log(1+\delta_3)$ in the three redshift bins, yielding the following values: $\rho_{0.4-0.7} = 0.21$, $\rho_{0.7-0.85} = 0.14$ and $\rho_{0.85-1} = 0.04$. Taking into account our sample size in each bin we estimate a significance level of $\sim2\sigma$ for $\rho_{0.4-0.7}$ and $\rho_{0.7-0.85}$ and $0.5\sigma$ for $\rho_{0.85-1}$, i.e., colour and density are uncorrelated at $z>0.85$ for $\log~M_\ast > 10.25$. 
This is consistent with the results of \citet{Coo06,Coo07}, who find a strong correlation between red fraction and local density, with the slope of the relation decreasing with look back time. The significance of the $f_{red}$-density relation in their colour limited sample is however higher ($>3\sigma$) than the significance of the relation we find for our stellar mass limited sample. \citet{Ger07} find a similar result for galaxy groups in the DEEP2 survey, such that galaxy groups have lower fractions of blue galaxies than the field. However, this difference is seen up to a higher redshift ($z\sim1.3$) than colour-density relation in our sample. Our result is also consistent with the findings of \citet{Cuc06} using the VIMOS VLT Deep Survey (VVDS) to construct a luminosity limited sample of 6582 galaxies and local densities measured on a larger scale ($5 h^{-1}$ Mpc) than the densities we use here. \citet{Cuc06} find that a clear relation between colour and local density at $0.25 < z <0.6$, which progressively disappears with higher redshift and is not present at $z\sim 0.9$, coinciding roughly with the redshift at which the colour-density relation disappears in our sample ($z>0.85$).

Analogous to the relation between $(U-B)$ and $(1+\delta_3)$ we find a similar correlation between $(U-B)$ colour and number of neighbours $N_{\rm 1Mpc}$. The top right panel of Figure~\ref{fig6} shows colour and colour fractions as a function of $\log N_{\rm 1Mpc}$. This colour-density relation has a slightly lower significance than the colour-{\it over}density relation between $(U-B)$ and $(1+\delta_3)$ and is also most significant in the lower and intermediate redshift bin.
We also detect a weak correlation between colour and mean harmonic radius $R_H$, i.e., compactness of the structure. This relation is shown in the bottom left panel of Figure~\ref{fig6}. $R_H$ is related to the local density, since galaxies in a high density will most likely have small projected separations. However, a high $R_H$ can be also achieved with a low number of galaxies, i.e., at a low density. The colour-compactness relation is similar to the colour-density relation, showing more red galaxies in structures with small $R_H$, most significantly in the intermediate redshift bin. Its significance is, however, lower than that of the colour-density relation.  
The bottom right panel of Figure~\ref{fig6} shows colour and colour fractions as a function of the dimensionless crossing time $t_c~H_0$. The crossing time and colours are possibly correlated in the intermediate redshift bin, in a way that structures with longer crossing times are populated by bluer galaxies. The significance of the correlation between $t_c$ and $f_{blue}$ is $\sim1\sigma$.

%---------------------------begin Figure 7------------------------------------------ 
\begin{figure}
\includegraphics[width=0.477\textwidth]{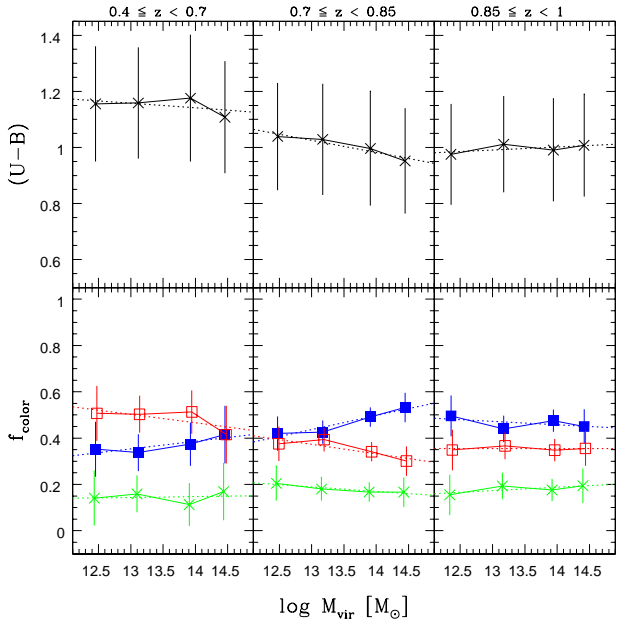}
\caption{Same as Figure~\ref{fig6} but for virial mass $\log~M_{\rm vir}$.
\label{fig7}}
\end{figure}
%-----------------------------end Figure 7 ----------------------------------------

\subsubsection{$(U-B)$ and $f_{colour}$ as a function of halo mass}

In the following the dependence between galaxy colour and parent dark matter halo mass is investigated. Only galaxies with spectroscopic redshifts are used to determine the virial mass of each galaxy's parent dark matter halo.
Figure~\ref{fig7} shows colour and colour fractions as a function of halo mass $\log~M_{\rm vir}$. In contrast to the colour-density relation there is no significant correlation between galaxy colour and $\log~M_{\rm vir}$. The correlation coefficients are $< 0.1$ with a significance of $<1\sigma$ at all redshifts. However, there is a trend that more massive haloes have slightly bluer central galaxies in the redshift range $0.7 \leq z < 0.85$. %The blue fraction in the most massive haloes ($\log~M_{\rm vir}\sim14.5$) is $f_{blue}=0.53\pm0.06$, while it is $f_{blue}=0.41\pm0.07$ at $\log~M_{\rm vir}\sim12.5$.
Additionally, $f_{blue}$ decreases faster with redshift at low $\log~M_{\rm vir}$ than at high $\log~M_{\rm vir}$. %At $\log~M_{\rm vir}\sim12.5$ the blue fraction decreases from $f_{blue} = 0.50\pm0.08$ at high $z$ to $f_{blue} = 0.35\pm0.11$ at low $z$. At the high-mass end ($\log~M_{\rm vir}\sim14.5$) $f_{blue}$ stays roughly constant between $f_{blue}=0.45\pm0.07$ (high $z$) and $f_{blue}=0.42\pm0.12$ (low $z$) with the rise at intermediate $z$ mentioned above.
This implies that galaxies in lower mass haloes become redder earlier. Alternatively high-mass haloes could accrete more blue field galaxies. However, this trend has only $\sim1\sigma$ significance. Note that the errorbars in the figure are the 1$\sigma$ rms deviation in each $M_{\rm vir}$ bin, not the measurement errors, which are negligible compared to the uncertainties due to the on average low number of objects around each galaxy. The virial mass is accurate within a factor of a few (3-5) due to the low numbers of objects we use to determine it ($\langle N_{\rm 1Mpc} \rangle \sim 4$).

\subsection{Rest-frame colours and colour fractions correlated with stellar mass}\label{colour-mass}

In analogy to the relation between colour and density and colour and halo mass, we now investigate the relation between colour and stellar mass $M_\ast$.  Since no environment measure sensitive to photometric redshift errors or edge effects is used in the following we use the full sample ($> 14,000$ galaxies) down to $\log~M_\ast = 10.25$ and $z = 1$.

%---------------------------begin Figure 8------------------------------------------
\begin{figure}
\includegraphics[width=0.477\textwidth]{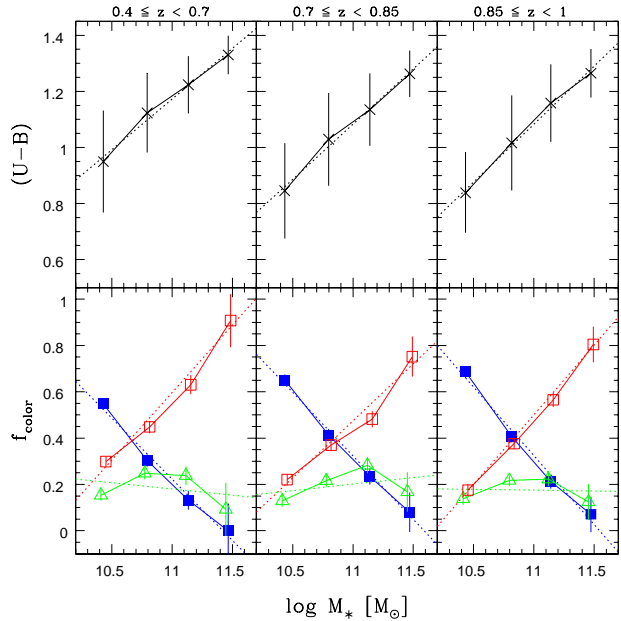}
\caption{Same as Figure~\ref{fig6} but for stellar mass $\log~M_\ast$.
\label{fig8}}
\end{figure}
%-----------------------------end Figure 8 ----------------------------------------

Figure~\ref{fig8} shows $(U-B)$ colour and colour fraction as a function of stellar mass ($\log~M_\ast$). As can immediately be seen, both colour and colour fractions depend strongly on galaxy stellar mass. The fraction of red galaxies $f_{red}$ changes with $M_\ast$ from $\sim$0.2 at $\log~M_\ast \sim10.5$ to $\sim$0.8 at $\log~M_\ast \sim 11.5$ at $z\sim1$. The $f_{colour}$-stellar mass relation is roughly constant with redshift, flattening slightly at the low-mass end from $f_{red} \sim0.2$ at $z\sim1$ to $f_{red} \sim0.3$ at $z\sim0.5$.
The strong relation between rest-frame colour and a galaxy's stellar mass is the basis of the relation between colour and mass-to-light ratio M/L found by \citet{Bel01}, which allows stellar masses to be determined from luminosity and rest-frame colours.
We compute the significance of the correlation of $f_{blue}$ and $f_{red}$ with stellar mass using the Spearman rank coefficient and its z-score, as already used above for the colour-density and colour-halo mass correlations. The correlation coefficient is $\rho \sim 0.4$ at all redshifts, corresponding to a significance level of $> 8 \sigma$.

Interestingly there are significantly more green valley galaxies at intermediate $M_\ast$ ($\log~M_\ast \sim 11$) than in the lowest and highest stellar mass bin at all redshifts. The green fraction changes from $f_{green} \sim 0.25$ at $\log~M_\ast \sim 11$ to $f_{green} \sim 0.15$ at $\log~M_\ast \sim 10.5$ and $\log~M_\ast \sim 11.5$. This difference has a $> 2 \sigma$ significance and is roughly constant at all redshifts. This might indicate that the transition from blue to red between $0.4 \leq z < 1$ is happening in intermediate mass galaxies, while it is mostly finished at high stellar masses and has yet to start at low stellar masses.

Our findings are roughly consistent with the results of \citet{Iov10} using the zCOSMOS group catalogue of \citet{Kno09} reaching up to $z \sim 0.8$. \citet{Iov10} find that the fraction of blue galaxies reaches `saturation' at the extremes of the stellar mass distribution, i.e. $F_{blue}\sim 1$ at low stellar mass ($\log~M^\ast \sim 10$) and $F_{blue}\sim 0$ at high stellar mass ($\log~M^\ast > 10.6$). At intermediate stellar masses the blue fraction varies with environment, being lowest in the group sample and highest in the sample of isolated galaxies. The difference becomes progressively larger at lower redshifts and is most distinctive at $0.25 < z < 0.45$. This trend seems to continue towards low redshift, where \citet{Kau03b}, using the SDSS, find that the fraction of galaxies with old stellar populations is rapidly increasing at stellar masses above $\log~M_\ast = 10.3$.

%---------------------------begin Figure 9------------------------------------------
\begin{figure*}
\includegraphics[width=0.32\textwidth]{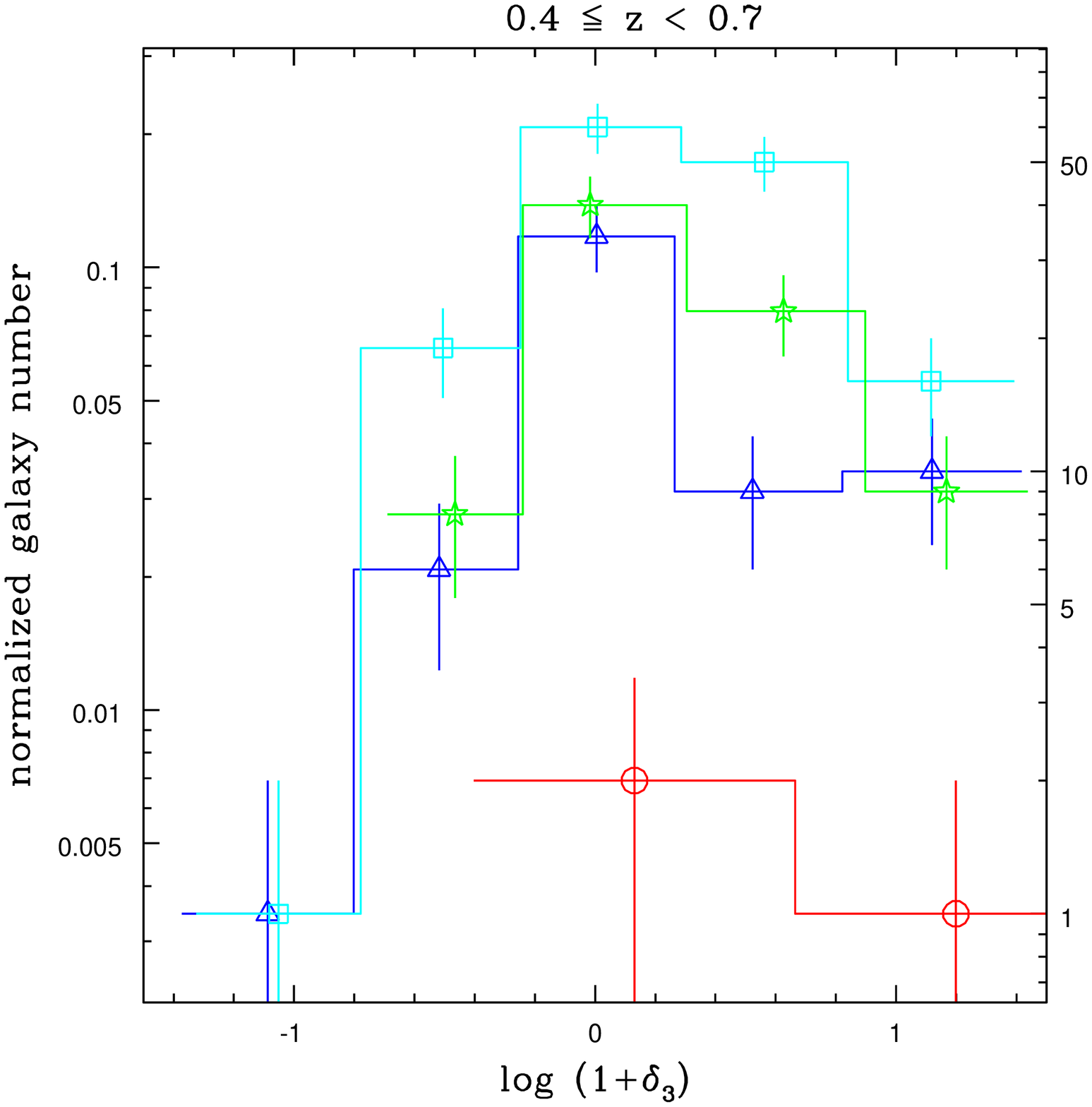}
\includegraphics[width=0.32\textwidth]{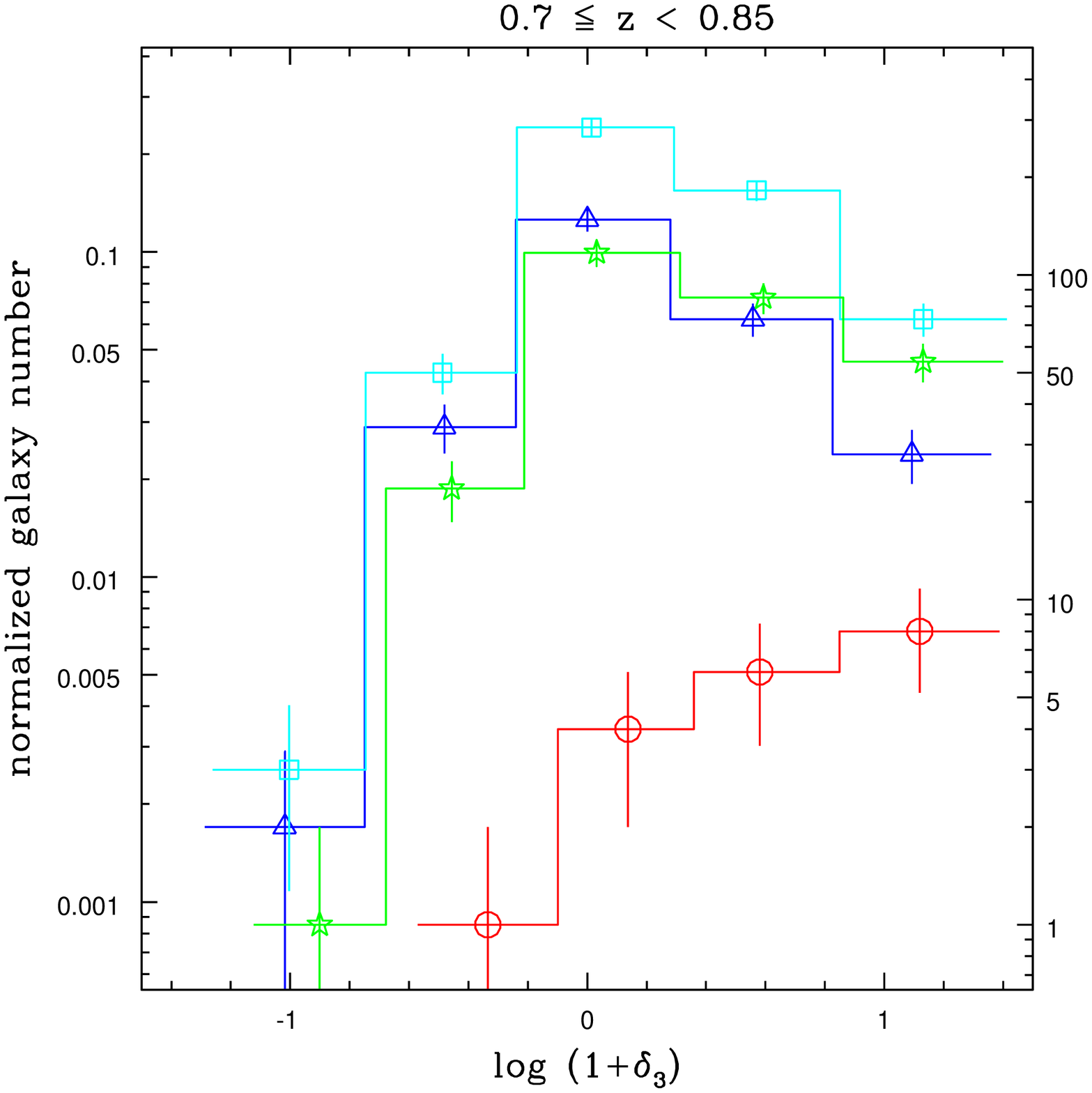}
\includegraphics[width=0.32\textwidth]{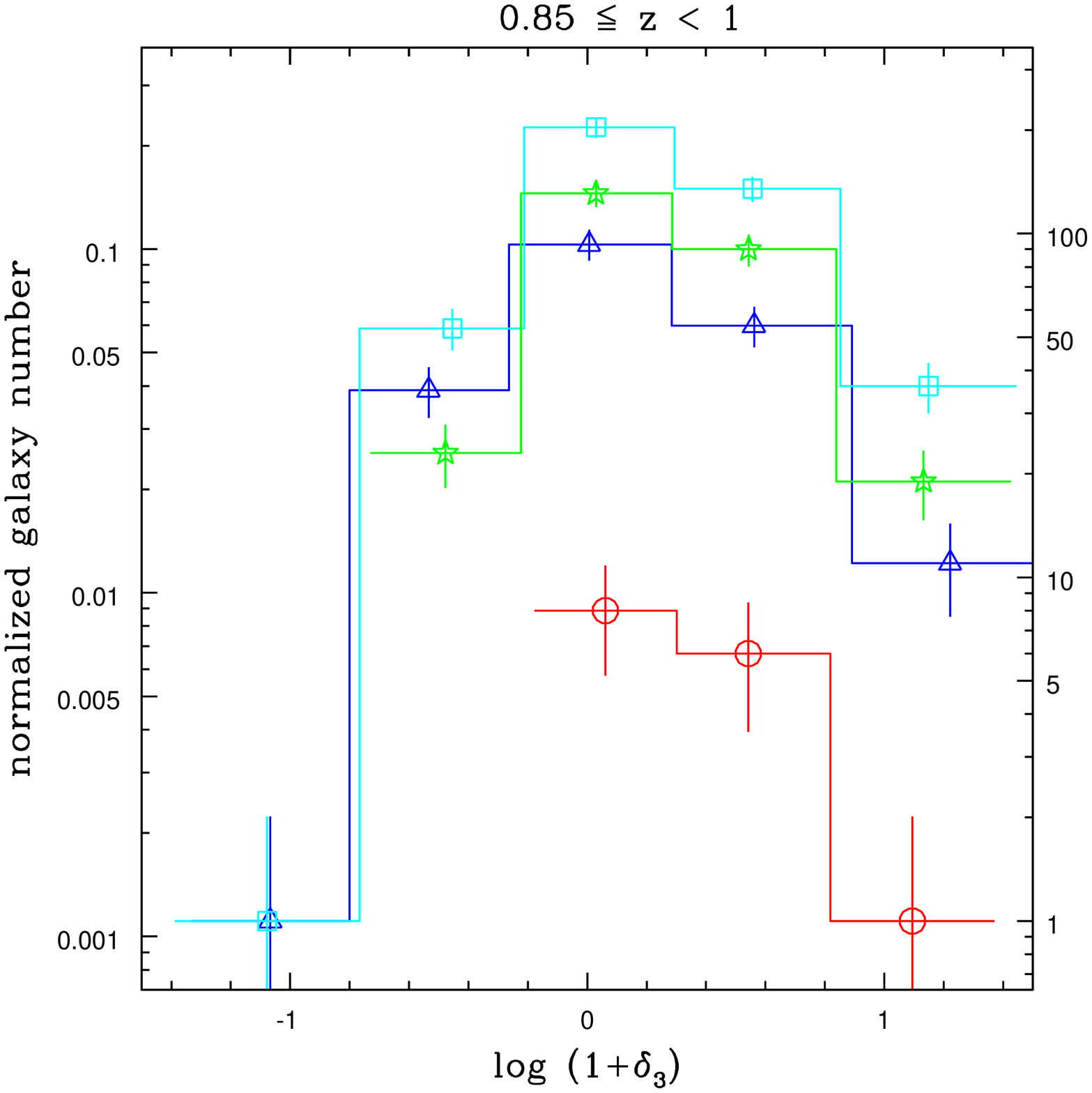}
\caption{Number of galaxies in the four stellar mass bins at different local densities. The stellar mass bins are colour coded and plotted with different symbols as in Figures~\ref{fig10} and \ref{fig11}: blue triangles ($\log~M_\ast < 10.5$), cyan squares ($10.5 < \log~M_\ast < 11$), green stars ($11 < \log~M_\ast < 11.5$), and red circles ($\log~M_\ast > 11.5$).
The scale of the y-axis on the left side shows the number of galaxies in each stellar mass bin and local density bin, normalized by the total number of galaxies in the respective redshift bin. For comparison reasons we show the number of galaxies in each bin without this normalization on the right side of each plot.
\label{fig9}}
\end{figure*}
%---------------------------end Figure 9------------------------------------------

\subsection{Disentangling dependencies on environment from dependencies on stellar mass}

\begin{figure}
\includegraphics[width=0.477\textwidth]{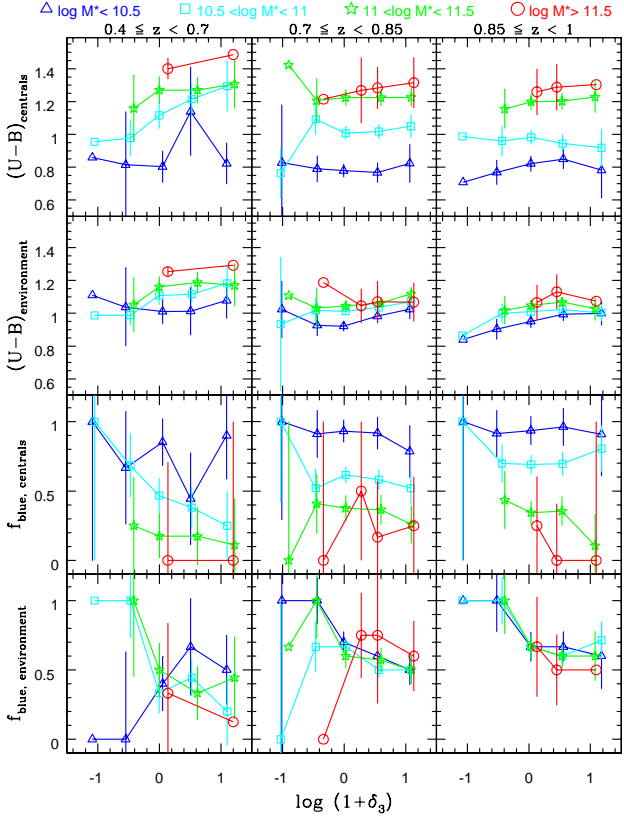}
\caption{$(U-B)$ colour and the blue fraction of galaxies ($f_{blue,centrals}$) and their environment ($f_{blue,environment}$) in different stellar mass bins: $(U-B)$ and $f_{blue}$ as a function of nearest neighbour density $\log~(1+\delta_3)$ for galaxies in four stellar mass bins. From top to bottom: $(U-B)$ of galaxies, mean $(U-B)$ of the galaxies' neighbours, $f_{blue}$ of galaxies,  $f_{blue}$ of neighbours. The stellar mass bins are colour coded and plotted with different symbols: blue triangles ($\log~M_\ast < 10.5$), cyan squares ($10.5 < \log~M_\ast < 11$), green stars ($11 < \log~M_\ast < 11.5$), and red circles ($\log~M_\ast > 11.5$). The errorbars represent the $3\sigma$ error of the mean in each bin.
\label{fig10}}
\end{figure}

\begin{figure}
\includegraphics[width=0.477\textwidth]{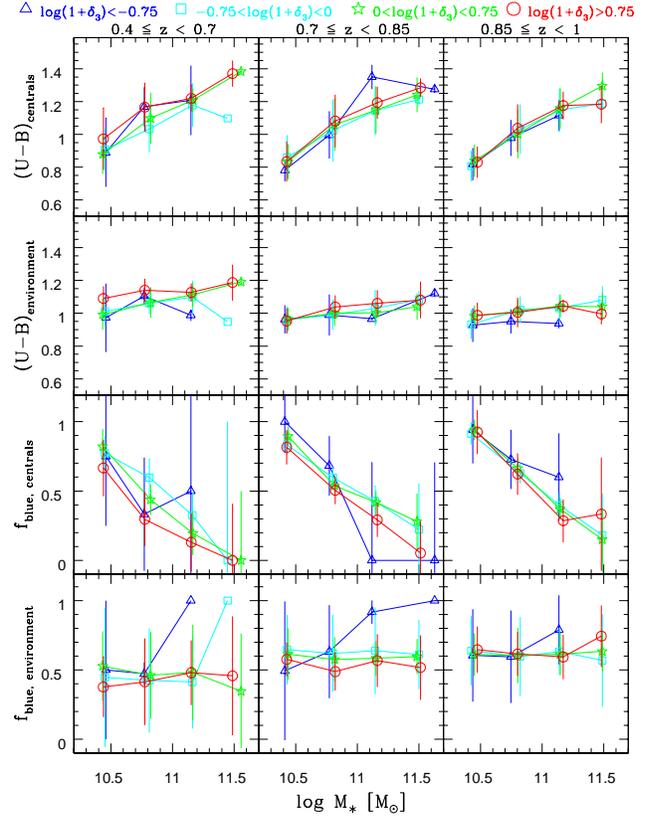}
\caption{$(U-B)$ colour and the blue fraction of galaxies ($f_{blue,centrals}$) and their environment ($f_{blue,environment}$) in different local density bins: $(U-B)$ and $f_{blue}$ as a function of stellar mass $\log~M_\ast$ for galaxies in four local density bins. From top to bottom: $(U-B)$ of galaxies, mean $(U-B)$ of the galaxies' neighbours, $f_{blue}$ of galaxies,  $f_{blue}$ of neighbours. The local density bins are colour coded and plotted with different symbols: blue triangles ($\log~(1+\delta_3) < -0.75$), cyan squares ($-0.75 < \log~(1+\delta_3) < 0$), green stars ($0 < \log~(1+\delta_3) < 0.75$), and red circles ($\log~(1+\delta_3) > 0.75$). The errorbars represent the $3\sigma$ error of the mean in each bin.
\label{fig11}}
\end{figure}

To disentangle the effects of local density and stellar mass on galaxy colours and colour fractions we split the sample into different bins of stellar mass and density (see Figures~\ref{fig9},~\ref{fig10} and \ref{fig11}). The stellar mass bins are the same as the bins used in Figure~\ref{fig4} and Figure~\ref{fig5}: red circles correspond to galaxies with $\log~M_\ast > 11.5$, green stars are $11.5 > \log~M_\ast > 11$, cyan squares comprise of galaxies between $11 > \log~M_\ast > 10.5$ and blue triangles are all galaxies below $\log~M_\ast < 10.5$, down to the limiting stellar mass of $\log~M_\ast = 10.25$. 
Analogous to the four stellar mass bins the sample is also divided into four local density bins defined in the following way: red circles are the densest environments with $\log~(1+\delta_3) > 0.75$, green stars are slightly overdense environments at $0.75 > \log~(1+\delta_3) > 0$, cyan boxes are for slightly underdense with $0 > \log~(1+\delta_3) > -0.75$ and blue triangles are the most underdense environments where $\log~(1+\delta_3) < -0.75$.
Figure~\ref{fig9} shows the numbers of galaxies in each stellar mass and local density bin. To compare the three redshift ranges we normalize the number of galaxies in each bin by the total number of galaxies in the respective redshift range. The four stellar mass bins are colour coded in the figure as described above, while the density is plotted on the x-axis.
This Figure shows at which local densities most galaxies of a given stellar mass are found and how this might change with redshift. We find that at intermediate redshifts there are more massive galaxies at higher densities. This redshift range also shows the strongest colour-density relation (see Figure~\ref{fig6}). 

\citet{Con07} discuss the evolution of the number densities of the most massive galaxies ($\log M_\ast > 11.5$) in our sample. They show that $\log M_\ast > 11.5$ galaxies are already in place by $z\sim1$ and their number densities do not evolve significantly after that. In Figure~\ref{fig9} we still see some evolution in the numbers of the most massive galaxies ($\log M_\ast > 11.5$, red circles) between $z\sim1$ and $z\sim0.8$. This evolution in taking place at the highest local densities. The same is seen for galaxies in the next stellar mass bin ($11.5 > \log M_\ast > 11$), which increase in numbers only at the highest relative overdensities. The lowest redshift bin does not have enough volume to probe the number densities of the most massive galaxies, due to their intrinsic rareness. Intermediate and low-mass galaxies ($\log M_\ast < 11$) also show increased numbers per unit volume at the highest local densities. The number densities of these galaxies are generally increasing between $z\sim1$ and $z\sim0.8$, but the increase is strongest in the most overdense environments.

We use the above stellar mass and local density bins to replot the overall colour density relation shown in Figure~\ref{fig6} and the overall colour-stellar mass relation (Figure~\ref{fig8}) in bins of stellar mass and relative density.
Figure~\ref{fig10} and Figure~\ref{fig11} then show the same plots as in Figure~\ref{fig6} and Figure~\ref{fig8} but with the data split into bins of relative density and stellar mass. For clarity only the blue fraction is plotted here. 
In Figure~\ref{fig10} and Figure~\ref{fig11} we also show the colour and blue fraction of the environment, i.e., mean $(U-B)$ of neighbours and fraction of blue galaxies in each galaxy's environment, denoted $(U-B)_{environment}$ and $f_{blue, environment}$ in contrast to $(U-B)_{centrals}$ and $f_{blue, centrals}$ for colours and blue fractions of individual galaxies. Note again that in this context `centrals' does not imply that these galaxies are indeed the central galaxy of their parent dark matter halo, but is used to distinguish between values for individual galaxies and mean values of neighbouring galaxies.

%---------------------------begin Figure 10------------------------------------------
%-----------------------------end Figure 10 ----------------------------------------

%---------------------------begin Figure 11------------------------------------------
%-----------------------------end Figure 11 ----------------------------------------

\subsubsection{The colour-density relation in different stellar mass bins}

Figure~\ref{fig10} shows how $(U-B)$ colour and the blue fraction ($f_{blue}$) of galaxies with different stellar masses depend on local density $\log~(1+\delta_3)$. The top row shows the $(U-B)$ colour of `central' galaxies, the second row the mean $(U-B)$ of the galaxies' environment as already used in Section~\ref{colour-z}.
The dependence of colour on density shown in Figure~\ref{fig6} largely disappears for low stellar mass ($\log~M_\ast < 10.5$) and high stellar mass ($\log~M_\ast > 11$) galaxies. However, there is a residual colour-density-relation for the intermediate mass galaxies (cyan, $11 >$ $\log~M_\ast > 10.5$). This relation is only present in the lowest redshift bin and disappears completely at $z \sim 0.8$. The same is seen for $f_{blue}$. The strong dependency of colour on stellar mass becomes clear in the plot: the different stellar mass bins are clearly separated in colour. The mean colour of the environment shows the overall colour-density relation and a slight dependence on stellar mass of the `central' galaxy: the most massive galaxies are surrounded by redder galaxies. This trend is weak, but becomes more evident at lower redshift, where the separation of the different stellar mass bins becomes clearer.

\subsubsection{The colour-stellar mass relation in different local density bins}

Figure~\ref{fig11} shows the colour-stellar mass relation in four bins of local density. The colour-stellar mass relation is clearly present in all density bins, but there is almost no colour separation between the different densities, reflecting the disappearance of the colour-density relation described above.
The colour-stellar mass relation is very similar at all densities, i.e. independent of local density, especially at high $z$. At low redshift the highest density bin seems to have a lower blue fraction, however the dispersion is large due to the small number statistics. 
The mean colour of a galaxy's environment $(U-B)_{environment}$ is also influenced by the stellar mass of a galaxy, such that more massive galaxies have on average slightly redder companions. This dependence however is not nearly as strong as the colour-stellar mass relation for individual galaxies ($(U-B)_{centrals}$). The colour difference between $\log~M_\ast \sim 10.5$ and 11.5 is $\Delta (U-B)_{centrals} = 0.4-0.5$ magnitudes, whereas the difference in the environment colour over the same stellar mass range is $\Delta (U-B)_{environment} = 0.15$ magnitudes.

\subsubsection{Correlations between density and stellar mass}

\begin{figure}
\includegraphics[width=0.35\textwidth, angle=270]{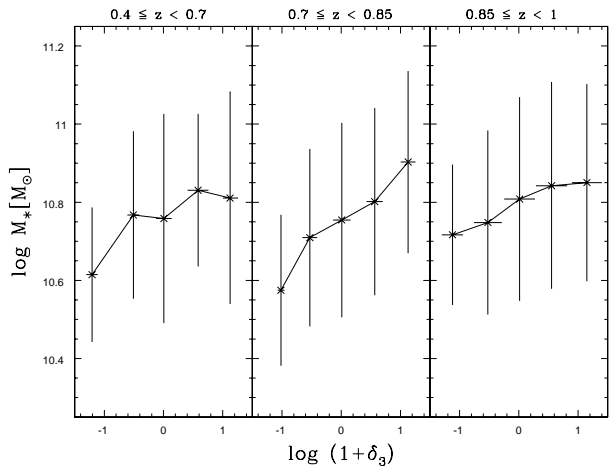}
\caption{Comparison between relative overdensity $\log~(1+\delta_3)$ and stellar mass $\log~M_\ast$ in the three redshift bins. The errorbars represent the $1\sigma$ spread in each bin.
\label{fig12}}
\end{figure}

Above we discussed the relations between galaxy colour and density on one hand and stellar mass on the other as two independent trends. However, how do stellar mass and local density correlate? Figure~\ref{fig12} shows the dependence of $M_\ast$ on $\log~(1+\delta_3)$. There is a trend that galaxies in higher local densities tend to have higher stellar masses, however this trend is weak and is only significant in the intermediate redshift bin, with a correlation coefficient $\rho_{0.7-0.85}=0.2$ at a significance of $\sim3\sigma$. This redshift bin also shows the strongest colour-density relation. In the other two redshift bins the correlation has a significance of $1.2\sigma$ ($0.4 \leq z < 0.7$) and $1.5\sigma$ ($0.85 \leq z < 1$).

\begin{figure*}
\includegraphics[width=0.485\textwidth]{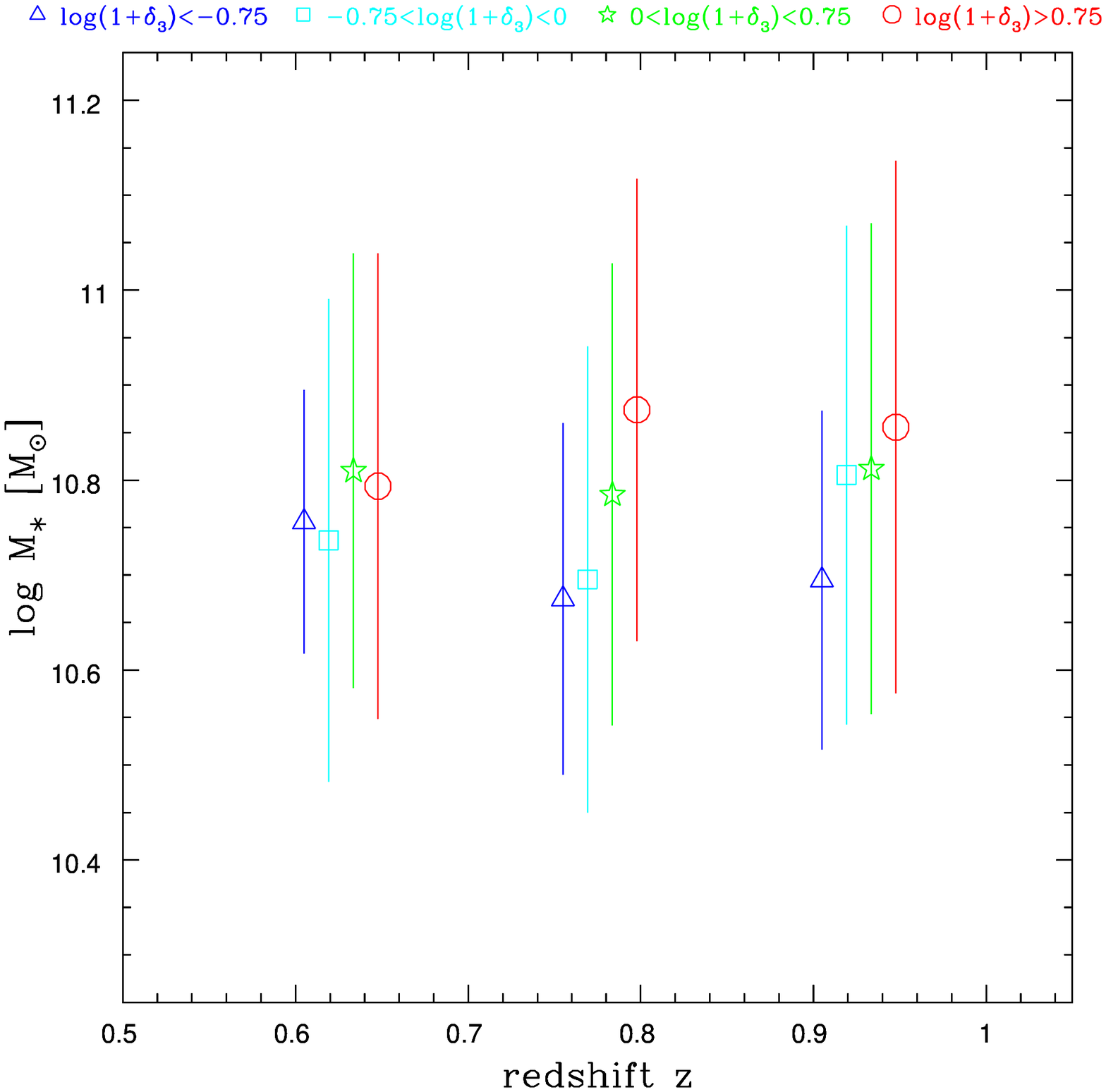}
\includegraphics[width=0.485\textwidth]{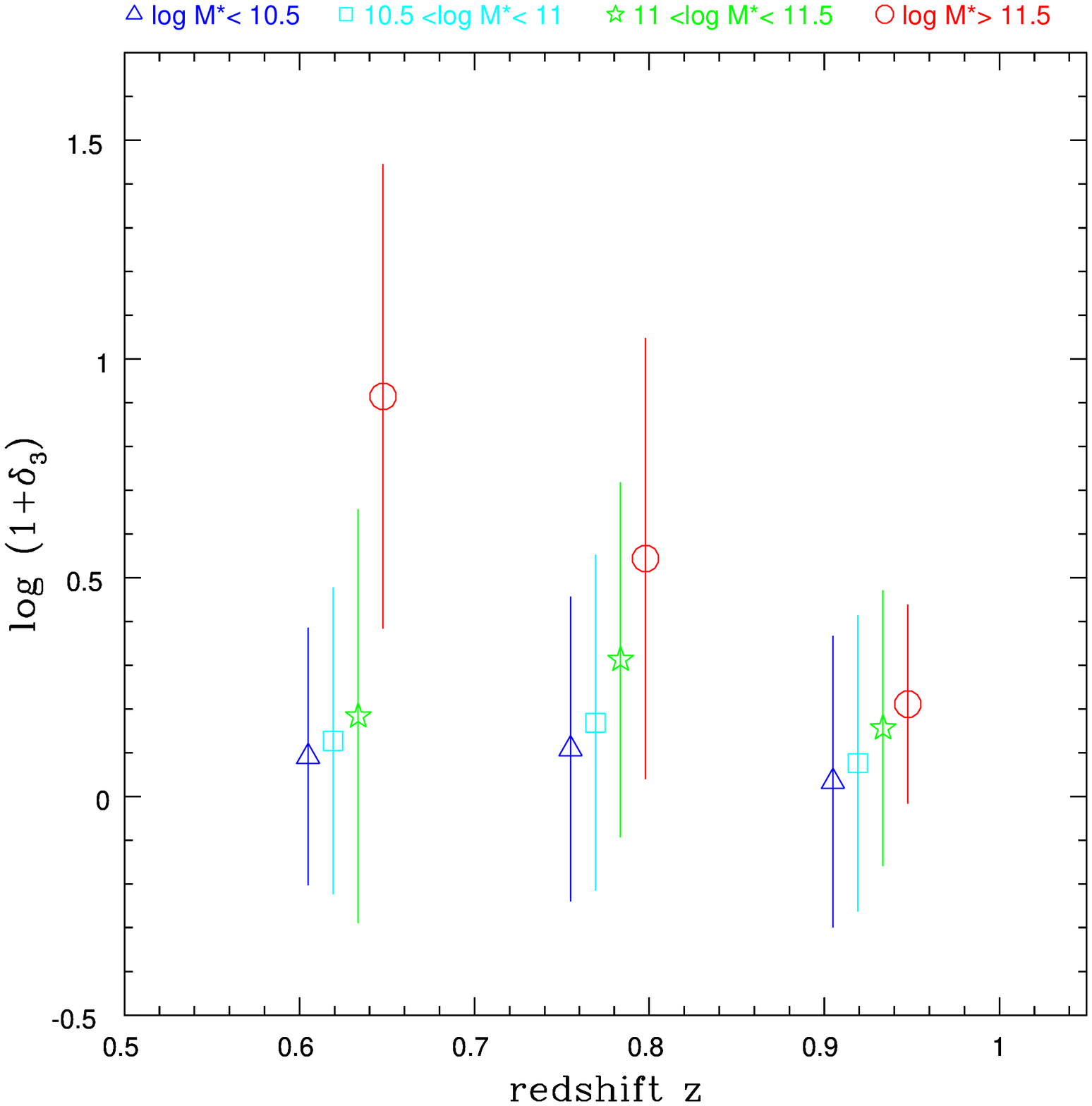}
\caption{Evolution of stellar mass and relative over-density with redshift. Left panel: evolution of stellar mass in different density bins. Right panel: evolution of densities in different stellar mass bins. The relative over-density of the most massive galaxies increases with redshift. The errorbars represent the $1\sigma$ dispersion in each bin.
\label{fig13}}
\end{figure*}

Is the slight redshift evolution of the density-stellar mass relation shown in Figure~\ref{fig12} due to an evolution of stellar mass at a given density or a change in density at a given stellar mass?
Figure~\ref{fig13} shows the redshift evolution of stellar mass at different local densities (left panel) and the redshift evolution of local density at different stellar masses (right panel). The weak correlation between stellar mass and density is visible, however, the stellar mass at a given density does not evolve with time.
The same is true for the evolution of densities at a given stellar mass, apart from the most massive galaxies which show a strong evolution of relative overdensity with time. Their environments change from about average densities to more than ten times the average, overdense regions. The same increase is seen in the number of neighbours $N_{\rm 1Mpc}$ as discussed earlier (see Figure~\ref{fig4}). This can be explained in terms of dynamical friction: since massive galaxies are more likely to be at the centre of their structure, neighbours are more likely to accumulate in their environment, increasing their local density over time. In our sample we indeed see evidence for the massive galaxies to be more likely located in the centre of mass and velocity of a galaxy association (see discussion of the $R$ parameter in Section~\ref{Rparam}).

%---------------------------begin Figure 12------------------------------------------
%-----------------------------end Figure 12 ----------------------------------------

%---------------------------begin Figure 13------------------------------------------
%-----------------------------end Figure 13 ----------------------------------------

\section{Discussion}\label{discussion}

In the following we discuss the implications of the results presented above and compare them with results from the literature.

\subsection{Environments of galaxies with different stellar masses}

Overall, we do not find a relation between galaxy stellar mass and radial velocity dispersion of galaxies within the galaxy's environment, or mass of the dark matter halo (as measured by $M_{\rm vir}$). This may partly be caused by the high uncertainties in measuring virial masses, due to the intrinsic low number of galaxies in the structures we probe here, possible interlopers or the possibility that some structures are unvirialized. However, as illustrated in Figure~\ref{fig2}, the velocity dispersions we measure are tracing halo masses in the Millennium Simulation down to $\log~M_{DM}\sim 12.5$. The lack of a (strong) correlation between stellar mass and virial mass is a challenge for models of hierarchical structure formation, since they expect more massive galaxies to be located in more massive structures (or parent dark matter haloes).

The environmental parameters do not change significantly with redshift, apart from $R_H$, which decreases with redshift at all stellar masses, and $R$, $N_{\rm 1Mpc}$ and $(1+\delta_3)$ for the most massive galaxies ($\log M_\ast > 11.5$). We find that the most massive galaxies tend to be more likely central galaxies (smaller $R$-parameter). They also move closer to the centre of mass and velocity of the structure, as suggested by decreasing $R$ with decreasing redshift. The most massive galaxies are located in higher absolute local densities ($N_{\rm 1Mpc}$) as well as in higher relative overdensities $(1+\delta_3)$ and accrete more galaxies over time, both in absolute numbers ($N_{\rm 1Mpc}$) and relative to the average ($(1+\delta_3)$). This can be explained by dynamical friction and is in general agreement with the expectations of hierarchical structure formation. 
The increase in local densities and number of companions might increase the rate at which merging is happening in these structures.
\citet{Con07,Con09} measure the merger rates for galaxies in our sample from morphological parameters and do not find an increasing merger rate with redshift for massive galaxies. 
The fraction of distorted ellipticals, however, is increasing in the same redshift range, reaching a maximum of about 30\% at $z \sim 0.7$ \citep{Con07}, which could be connected to the increasing local densities.
% The merger rates for galaxies in the same sample were measured by \citet{Con07,Con09}. They find that the merger rates of galaxies at $\log~M_\ast>11$ determined from CAS parameters \citep[Concentration, Asymmetry, clumpinesS,][]{Con03} are roughly constant or possibly declining from $z \sim 1$ to $z \sim 0.4$, the redshift range we study here. The fraction of distorted ellipticals, however, is increasing in the same redshift range, reaching a maximum of about 30\% at $z \sim 0.7$ \citep{Con07}. The distorted morphologies could be connected to the higher local densities. However, we observe the rise in local densities only for the most massive galaxies with stellar masses $\log~M_\ast > 11.5$ which make up only about 6\% of galaxies more massive than $\log~M_\ast = 11$. This could also explain why we do not see an increase in the merger rates, as the high local densities of $\log~M_\ast > 11.5$ galaxies would suggest. An other possible difference are the time-scales traced by CAS-derived merger rates compared to high local densities, which are rather anticipating future events than recent merging.

Some of the environmental parameters are connected to intrinsic galaxy properties of colour and stellar mass: we find a correlation between colour and local density, overdensity and compactness, as well as between local density, compactness and stellar mass, as discussed in the following.

\subsection{Relations between galaxy colour and stellar mass, local density and halo mass}

There is a very strong correlation between galaxy colour and stellar mass ($> 8\sigma$ significance) at all redshifts. This colour-stellar mass relation evolves only slightly with redshift. The stellar mass where red and blue fractions cross over, i.e., where there is an equal fraction of blue and red galaxies, occurs on average at $\log M_\ast \sim 10.8$ and evolves slightly from $\log M_\ast \sim 10.9$ at high $z$ to $\log M_\ast \sim 10.7$ at low $z$. In the local universe this transition occurs at lower stellar masses: using the SDSS \citet{Kau03a,Kau03b} find a sharp transition from the dominance of blue to red galaxies at $\log M_\ast \sim 10.3$.
Our results also agree with the findings of a recent study by \citet{Iov10} who investigate the relationship between colour, stellar mass and environment up to $z \sim 0.8$. They find a qualitatively similar strong dependence of the blue fraction on stellar mass in all environments studied (groups, field and isolated galaxies). 

We find a weak correlation between colour and local density ($\sim 2 \sigma$ significance) for the whole sample and also a weak correlation between density and stellar mass ($\sim3\sigma$ significance). The colour-density and density-stellar mass relations change in a similar way with redshift: both relations are most significant at intermediate redshifts, and they both disappear at high $z$. 
This is in good agreement with previous similar studies \citep{Cuc06,Coo06,Coo07,Cas07}. \citep{Cuc06} find a similar progressive disappearance of the colour-density relation with redshift, even though they use galaxy densities measured on much larger scales of 5 $h^{-1}$ Mpc. \citep{Coo07} however find that a strong relation between colour and local density persists out to $z>1$. This might partly be caused by their sample selection, which is rest-frame $B$-band luminosity limited. We cannot confirm the colour-density relation at $z>0.85$ in our $K$-band selected sample.

If the sample is split up in different stellar mass bins, the colour-density relation disappears completely for lowest and highest mass galaxies, remaining only in the stellar mass range $10.5 < \log~M_\ast < 11$, although at a lower significance than for the whole sample. We conclude that the overall colour-density relation is powered by these intermediate mass galaxies, as well as by the fact that there are more low-mass galaxies at lower densities and vice versa. Stellar mass and relative overdensity  are not strongly correlated, however, the correlation accounts for most of the colour-density relation seen in our sample. In other words, the colour-density relation is a combination of a strong colour-stellar mass relation and a weak stellar mass-density relation. The strong colour-stellar mass relation is present at all local densities and it does not vary with density, apart from possibly in the lowest redshift bin of $0.4 < z < 0.7$. The colours of galaxies at a given stellar mass are very similar at all local densities. 
These results are supporting the evidence emerging from previous studies using stellar mass limited samples, suggesting that the density dependence of galaxy colours is regulated mainly by stellar mass. The relative numbers of red and blue galaxies depend on the environment \citep[e.g.][]{Cas07}, leading to the observed colour-density relation. The variations in relative numbers of red and blue galaxies can be explained in terms of variations of the stellar mass function with environment, which are present as early as $z \sim 1$ and are becoming stronger towards lower redshift \citep[e.g.][]{Bun06,Bol09}.
\citet{Tas09} (and similarly \citet{Iov10}) suggest the presence of a limiting stellar mass ($\log~M^\ast \sim 10.6$) above which a galaxy's shape and colour is mainly correlated to its stellar mass rather than its environment.
\citet{vdW07} study the morphology-density relation at $0.6 < z < 1$, using 207 galaxies from the Chandra Deep Field South (CDF-S) down to $\log~M_\ast = 10.6$ and compare them to a similar mass selected sample from the SDSS containing 2003 galaxies at $0.020 < z < 0.045$. They find a strong morphology-density relation for massive galaxies which has remained constant since at least $z ~ 0.8$. However, they detect an increasing early-type fraction only above absolute densities of $\sim 10$ Mpc$^{-2}$ (their highest density `field' data point) and in the two high density points representing the cluster environment. This might explain why we detect a relatively weak density dependence, since the DEEP2 survey does not cover the high density cluster environment \citep[e.g.][]{Ger07}.

The intermediate mass galaxies, which still show a colour-density relation, also show the most colour evolution with redshift, reddening substantially between $z\sim1$ and $z\sim0.4$. Interestingly, they also have the highest fraction of green valley galaxies, i.e., galaxies possibly in transition from the blue cloud to the red sequence. This suggests that in the redshift range we observe those intermediate mass galaxies are transiting from blue to red in environments of higher local densities.
This is consistent with the results of a recent study of blue fractions in galaxy groups up to $z \sim 0.8$ \citep{Iov10}, who find that blue fractions of intermediate mass galaxies show an environmental dependence which is not seem for galaxies with higher and lower stellar mass.

The mean colour of a galaxy's environment is as well sensitive to the galaxy's stellar mass: the most massive galaxies are surrounded by redder galaxies. The difference in the mean environment colour $(U-B)_{environment}$ is however small with only 0.1-0.2 magnitudes between the lowest and the highest mass galaxies.

We also find a correlation between colour and compactness of the group structure ($R_H$) and perhaps between colour and crossing times $t_c$, but only in the redshift range $0.7 \leq z < 0.85$. The colour-compactness relation has a slightly lower statistical significance than the colour-density relation, but is still significant at $\sim 2\sigma$ in the redshift range $0.7 \leq z < 0.85$. Since $R_H$ can be small also at low densities, this suggests that galaxy colour may be influenced by the presence of a close neighbour regardless of the local density.

There is no significant correlation between galaxy colour and parent dark matter halo mass in our redshift range, however, there is a trend such that the most massive haloes are populated by bluer galaxies at intermediate redshift ($0.7<z<0.85$). This is different to what was found at lower redshifts. \citet{Wei06} investigate SDSS group galaxies in a redshift range $0.01 \leq z \leq 0.2$ and find a weak correlation between colour and halo mass: high-mass haloes are populated by redder galaxies. 
However, \citet{Wei06} use a different method to estimate the total group mass (or halo mass): they use the integrated group light as a halo mass estimator. Since there is a correlation between galaxy colour and galaxy luminosity, this approach may automatically lead to a correlation between galaxy colour and group halo mass. \citet{Bam09} investigated the relation between morphology and group halo mass in the Galaxy Zoo sample and found no correlation between morphology and group mass, measured either by group velocity dispersion or integrated light.
A correlation between colour and halo mass for {\it central} galaxies was found by \citet{Ski09}, who argue that central galaxies in more massive haloes tend to have redder colours. What we investigate here is somewhat different from the results of \citet{Ski09}: we investigate whether more massive haloes are inhabited by on average redder galaxies, taking into account all galaxies at the respective halo mass, not only the central galaxies. We do not find any significant differences between average colour and colour fractions of galaxies in haloes of different mass based on our virial mass measurements, apart from a weak trend at intermediate redshift that higher mass haloes have higher fractions of blue galaxies.

The fact that the relation between colour and stellar mass is much stronger than the relation between colour and environmental parameters might be partly due to the fact that stellar masses can be measured more accurately than environment. The expected error of the stellar masses is about 0.3 dex, or about a factor of 2, while the  uncertainty of the virial mass due to low galaxy numbers is a factor of 3-5, corresponding to 0.5-0.7 dex. The 1$\sigma$ error of the nearest neighbour densities, estimated by comparison with DEEP2 mock catalogues \citep{Coo06}, is 0.5 in units of $\log (1 +\delta_3)$, a factor of about 3, similar to the uncertainty in the virial mass. There is the possibility that the high uncertainty in virial mass washes out any weak trends between galaxy colour, colour fractions and halo mass. However, the uncertainties in virial mass are similar to the uncertainties in local density and we do detect an overall colour-density relation in our sample.

Another issue is the range of environments sampled by the POWIR/DEEP2 survey: the survey contains hardly any massive galaxy clusters, i.e. systems with more than 10 members and velocity dispersions above $\sigma \sim 700$ km s$^{-1}$ \citep{Ger07}. The lack of this type of environment certainly influences the measured colour-density relation in our sample, since some of the processes that are density dependent and that most efficiently influence galaxy colours, like ram pressure stripping or galaxy harassment are only occurring in very massive structures: ram pressure stripping requires the presence of high velocities and a high density intra cluster medium, both retained by a deep potential well of a high-mass structure. Equally galaxy harassment, defined as repeated high velocity encounters, can only occur in massive structures with a high velocity dispersion like galaxy clusters. However, one has to keep in mind, that galaxy clusters are not a very typical environment and that the evolution of an average galaxy does not involve the cluster environment. The lack of massive clusters in our sample is therefore not a significant bias for quantifying the environmental influence on the evolution of an average galaxy.

\subsection{Photo-z versus spec-z sample}

The stellar mass limited full sample used in this study consists of 14,563 galaxies, $\sim30\%$ of which (4101 galaxies) have spectroscopic redshifts. These numbers allow us to compare the results for the full and spec-z only sample and discuss the limitations of photometric redshifts in this context.

As discussed above the velocity related environmental characteristics are very sensitive to the inclusion of photometric redshifts, and need the accuracy of spectroscopic redshifts. The mean value of the velocity dispersion using the full sample is $\sim 400$ km s$^{-1}$, which is the result of the velocity interval within which a neighbour galaxy is included ($\Delta v = \pm 1000$ km s$^{-1}$). Even very accurate photometric redshifts have errors much larger than this velocity interval, errors even much larger than the velocity dispersion of massive galaxy clusters. Photometric redshifts with the currently available accuracy are therefore not useful for the calculation of dynamical properties of structures, such as velocity dispersion and virial mass. 

The density estimates are also sensitive to photo-z errors \citep{Coo05,Guz07}, although not as much as velocity related quantities. The use of photo-z's increases the number of interlopers and smears out information along the line of sight. 
%However, using a statistical correction and a statistically significant sample size minimizes this effect: normalizing by the median of the sample in slices of redshift comparable to (or larger than) the photo-z error is one way to deal with this problem. The relative overdensity $(1+\delta_3)$ is normalized by the median density of the sample in slices of $\Delta z = 0.04$ and should be less affected by the photo-z errors.
The comparison between the full and spec-z only sample shows that the relation between colour and $N_{\rm 1Mpc}$ changes when galaxies with photo-z's are included: the correlation is still there, but at a lower significance of only $\sim1\sigma$ (compared to $>2\sigma$ for spec-z's only). The same is seen for the correlation between colour and $R_H$: it becomes weaker, but not as much as the colour-density relation. Its significance drops from $\sim2\sigma$ to $\sim1.5\sigma$.

The intrinsic galaxy properties of colour and stellar mass are not as sensitive to the photo-z errors, since the random errors of the SED fitting and the photometry are already larger than the expected error introduced by using photometric redshifts \citep{Bun06}. The colour-stellar mass relation is very robust against inclusion of photo-z's.

Another issue is the incompleteness in the spectroscopic redshift sample. This sample is not stellar mass limited at $\log M_\ast = 10.25$. The completeness limit is redshift and colour dependent, due to the DEEP2 $R$-band magnitude cut-off at $R_{AB} < 24.1$, leading to the loss of faint red galaxies at high $z$. The correlations between colour, local densities, virial mass and stellar mass should not be affected by this bias however. The local densities are divided by the median in each $\Delta z = \pm 0.04$ redshift slice to account for the redshift dependent completeness limit. The colour bias could affect the colour-density relation at high $z$ if faint red galaxies at that redshift preferred a certain environment. However, since we do not see a colour-density relation in any stellar mass bin at high $z$ and low-mass red galaxies are not very numerous, this possible bias is unlikely to influence the overall colour-density relation. 
The colour-halo mass relation, based on the velocity dispersion, should also not be affected, since the velocity dispersion of a group is well sampled without including the complete group population and should not strongly depend on colour or mass of the sampled galaxies. \citet{Got05} found that red galaxy cluster members in the SDSS have lower velocity dispersions than blue cluster galaxies, although the difference is marginal, of the order of a few percent. 
The colour-stellar mass relation is not influenced by the spectroscopic redshift incompleteness, since in this case we use the complete sample, i.e., galaxies with spectroscopic and photometric redshifts.

\section{Summary and conclusions}\label{conclusion}

We investigate the correlations between galaxy environment, rest frame colour and stellar mass in a $K$-band selected, stellar mass limited sample with a high number of secure spectroscopic redshifts, and additional accurate photometric redshifts from the POWIR/DEEP2 surveys. The sample is complete down to a stellar mass of $\log~M_\ast = 10.25$ and up to a redshift of $z = 1$. Different characteristics of `galaxy environment' are included in this study, including local galaxy overdensity, number of neighbours within a fixed aperture, their crossing times and mean harmonic radius, as well as their velocity dispersion and virial mass as an indicator of the structures parent dark matter halo mass. We then use these quantities to disentangle the effects of stellar mass and environment on galaxy colour. 
In the following we summarize our most important findings:

\begin{enumerate}
\renewcommand{\theenumi}{\arabic{enumi}.}
\item  Average galaxy colour $(U-B)$ and the fraction of blue galaxies depends strongly on galaxy stellar mass (see Figures~\ref{fig8} and \ref{fig11}). \\

\item  There is a weak correlation of galaxy colour with local number density (see Figures~\ref{fig6} and \ref{fig10}). This correlation becomes weaker with higher redshift and is not detectable at $z\sim1$. \\

\item  There is a weak correlation between stellar mass and local number density, such that higher mass galaxies are preferentially found in higher density environments (see Figure~\ref{fig9} and Figure~\ref{fig12}). The stellar mass-density relation also disappears at $z\sim1$. This is due to the fact that relative local densities of the highest mass galaxies are decreasing with higher redshift, and not due to an evolution of stellar mass at the highest local densities (see Figure~\ref{fig13}). In other words, at a given (high) stellar mass the relative overdensity increases with time, while at a given density the stellar mass of a galaxy does not evolve.\\

\item  When the sample is split up in different stellar mass bins, the colour-density relation largely disappears in all but the intermediate mass bin ($10.5 <$ log~$M_\ast < 11$). The colour-stellar mass relation, however, is very strong at all local densities and is independent of local density (see Figure~\ref{fig10} and Figure~\ref{fig11}). The overall colour-density relation is then powered by the relation between colour and stellar mass, and stellar mass and local density (see Figure~\ref{fig12}). In other words, the colour-density relation is a combination of a strong colour-stellar mass relation and a weak stellar mass-density relation. \\

\item  Colour and local density are still related in intermediate mass galaxies ($10.5 <$ log~$M_\ast < 11$) at lower redshifts ($0.4 < z < 0.7$), possibly shifting towards higher stellar masses ($11 <$ log~$M_\ast < 11.5$) at higher redshifts ($0.8 < z \leq 1$, see Figure~\ref{fig10}). The colour fractions $f_{blue}$, $f_{green}$ and $f_{red}$ and the colour evolution with $z$ (Figures~\ref{fig5} and \ref{fig8}) suggest that these intermediate mass galaxies are moving onto the red sequence in higher density environments at redshifts of $z \sim 0.7$. By that redshift higher mass galaxies have mostly moved to the red sequence already, although there remains a fraction of blue galaxies of $\sim$ 20\% at $\log~M_\ast > 11$. 
Low-mass galaxies ($\log~M_\ast < 10.5$) show a very high fraction of blue galaxies at all redshifts, with very little dependence on density.\\

\item  There is no significant correlation of galaxy colour or blue fraction with halo mass (virial mass). However, we find that in low-mass haloes $f_{blue}$ might decrease with cosmic time, while it stays roughly constant in high-mass haloes (see Figure~\ref{fig7}). \\

\item  Apart from an increase of the local density around the most massive galaxies, the local environment of galaxies does not change significantly from $z\sim 1$ to $z \sim 0.4$. However, we do detect a trend for increasing compactness (smaller harmonic radius $R_H$) with decreasing redshift, which is most significant in galaxies with high stellar mass ($\log~M_\ast > 11$).
We have not detected a significant correlation between galaxy stellar mass and velocity dispersion or virial mass (see Figure~\ref{fig4}), which might be partly due to the difficulty of measuring those quantities in galaxy structures with such few members.
Most structures are already in place by $z\sim1$, and are not growing significantly after that, which might be expected in a universe dominated by a cosmological constant. \\

\end{enumerate}

\noindent This study covers the fairly well studied redshift range of $0.4 < z < 1$ \citep[][among others]{Smi05,Pos05,Cuc06,Coo06,Coo07,Cas07,vdW07,Tas09,Iov10}. Above a redshift of $z\sim1.5$ the environmental influence on galaxy properties is unclear. A study similar to the present one but up to a redshift of $z\sim3$ utilizing the GOODS-NICMOS Survey \citep[GNS, Conselice et al., in preparation,][]{Bui08,Blu09}, the deepest $H$-band selected survey up to date, will be carried out in a forthcoming paper (Gr\"utzbauch et al, submitted).

\section*{Acknowledgments}

The authors would like to thank Darren Croton for providing the mock light cone from the Millennium Simulation.
The Millennium Run simulation  was carried out by the Virgo Supercomputing Consortium at the Computing Centre of the Max-Planck Society in Garching. The semi-analytic galaxy catalogue is publicly available at http://www.mpa-garching.mpg.de/galform/agnpaper. 

The DEEP2 redshift survey is supported by the NSF grants AST00-71048, AST00-71198, AST05-07428, AST05-07483, AST08-07630 and AST08-08133. The Palomar and DEEP2 surveys would not have been completed without the active help of the staff at the Palomar and Keck observatories. We particularly thank Richard Ellis and Sandy Faber for advice and their participation in these surveys. 
We acknowledge funding to support this effort from a National Science Foundation Astronomy \& Astrophysics Fellowship and from the UK Science and Technology Facilities Council (STFC). Support for the ACS imaging of the EGS in GO programme 10134 was provided by NASA through NASA grant HST-G0-10134.13-A from the Space Telescope Science Institute, which is operated by the Association of Universities for Research in Astronomy, Inc., under NASA contract NAS 5-26555. 

J.V. acknowledges the School of Physics and Astronomy (University of Nottingham) for granting access to its facilities, as well as all the
members of the School for sharing with him a wonderful working and personal environment.

The authors also wish to recognize and acknowledge the very significant cultural role and reverence that the summit of Mauna Kea has always had within the indigenous Hawaiian community. We are most fortunate to have the opportunity to conduct observations from this mountain.

\end{document}